\documentclass[sigconf]{acmart}

\usepackage{packages}
\usepackage{layouts}

\acmYear{2022}\copyrightyear{2022}
\setcopyright{rightsretained}
\acmConference[IMC '22]{ACM Internet Measurement Conference}{October 25--27, 2022}{Nice, France}
\acmBooktitle{ACM Internet Measurement Conference (IMC '22), October 25--27, 2022, Nice, France}
\acmPrice{}
\acmDOI{10.1145/3517745.3561440}
\acmISBN{978-1-4503-9259-4/22/10}

\makeatletter
\newcommand\EatSpacesHack{\@bsphack\@esphack}
\makeatother
\iffalse
\newcommand\reviewfix[1]{{\color{red}\sffamily\bfseries [RF:
		#1]}\EatSpacesHack}
\else
\newcommand\reviewfix[1]{\EatSpacesHack}
\fi
\makeatother

\begin{document}

\title{Rusty Clusters? Dusting an IPv6 Research Foundation}

\author{Johannes Zirngibl}
\affiliation{%
	\institution{Technical University of Munich}
	\country{Germany}
}
\email{zirngibl@net.in.tum.de}

\author{Lion Steger}
\affiliation{%
	\institution{Technical University of Munich}
	\country{Germany}
}
\email{stegerl@net.in.tum.de}

\author{Patrick Sattler}
\affiliation{%
	\institution{Technical University of Munich}
	\country{Germany}
}
\email{sattler@net.in.tum.de}

\author{Oliver Gasser}
\affiliation{%
	\institution{Max Planck Institute for Informatics}
	\country{Germany}
}
\email{oliver.gasser@mpi-inf.mpg.de}

\author{Georg Carle}
\affiliation{%
	\institution{Technical University of Munich}
	\country{Germany}
}
\email{carle@net.in.tum.de}




\begin{abstract}
	The long-running \hitlist service is an important foundation for IPv6 measurement studies.
	It helps to overcome infeasible, complete address space scans by collecting valuable, unbiased IPv6 address candidates and regularly testing their responsiveness.
	However, the Internet itself is a quickly changing ecosystem that can affect long-running services, potentially inducing biases and obscurities into ongoing data collection means.
	Frequent analyses  but also updates are necessary to enable a valuable service to the community.

	In this paper, we show that the existing hitlist is highly impacted by the \acl{gfw}, and we offer a cleaned view on the development of responsive addresses.
	While the accumulated input shows an increasing bias towards some networks, the cleaned set of responsive addresses is well distributed and shows a steady increase.

	Although it is a best practice to remove aliased prefixes from IPv6 hitlists, we show that this also removes major content delivery networks.
	More than \sperc{98} of all IPv6 addresses announced by Fastly were labeled as aliased and Cloudflare prefixes hosting more than \sm{10} domains were excluded.
	Depending on the hitlist usage, \eg higher layer protocol scans, inclusion of addresses from these providers can be valuable.

	Lastly, we evaluate different new address candidate sources, including target generation algorithms to improve the coverage of the current \hitlist.
	We show that a combination of different methodologies is able to identify \sm{5.6} new, responsive addresses.
	This accounts for an increase by \sperc{174} and combined with the current \hitlist, we identify \sm{8.8} responsive addresses.
\end{abstract}

\maketitle

\section{Introduction}
\label{sec:introduction}
The usage and importance of IPv6 are steadily increasing \cite{ripev6growth,googlev6growth,apnicv6stats}.
With the IPv4 address depletion of all but one \ac{rir}, the necessity to deploy IPv6 is prevalent for more and more operators and content providers.
While this development is generally positive, it imposes fundamental difficulties to research and network analysis.
The immense size of the address space, combined with the sparse distribution of used addresses renders active IPv6 measurements difficult.
While tools like \zmap effectively scan the complete IPv4 address space, complete scans for IPv6 are impossible.

\begin{figure}
	\begin{tikzpicture}[
		entity/.style={draw, minimum width=4cm, minimum height=1cm},
		innerentity/.style={draw, minimum width=5cm,minimum height={height("Cap") + 3mm}, },
		inputentity/.style={draw, minimum width=3cm,minimum height={height("Cg") + 3mm}, },
		scaneentity/.style={minimum width=1.5cm},
		nameentity/.style={minimum width=2cm},
		dott/.style={minimum width=1.5cm},
		node distance=0.5cm
		]%

		\node[innerentity, rounded corners=6pt] (blocklist) {Blocklist Filter};
		\node[innerentity, rounded corners=6pt, below = of blocklist,color=TUMDarkGreen] (gfw) {Great Firewall of China Filter (\Cref{sec:hitlist})};
		\node[innerentity, rounded corners=6pt, below = of gfw] (apd) {Aliased Prefix Filter (\Cref{sec:aliased})};
		
		\node[minimum width=4cm,minimum height={height("Cap") + 3mm},below=of apd] (unresponsive) {};
		
		\node[draw, minimum width=2.1cm,minimum height={height("Cap") + 3mm}, rounded corners=6pt, left = -57 pt of unresponsive] (filter) {30-day Filter};
		
		\node[scaneentity, rounded corners=6pt, below = 25pt of unresponsive] (tcp80) {TCP/80};
		\node[scaneentity, rounded corners=6pt, right = 0pt of tcp80] (udp443) {UDP/443};
		\node[scaneentity, rounded corners=6pt, right = 0pt of udp443] (udp53) {UDP/53};
		
		\node[scaneentity, rounded corners=6pt, left = 0pt of tcp80] (tcp443) {TCP/443};
		\node[scaneentity, rounded corners=6pt, left = 0pt of tcp443] (icmp) {ICMP};

		\node[innerentity, rounded corners=6pt, below = 20pt of tcp80] (responsive) {Responsive Addresses};

		\node[entity, fit=(icmp) (udp53)] (zmap) {};
		
		\draw[-latex] ([shift={(-0.4,0)}]blocklist.south) -- ([shift={(-0.4,0)}]gfw.north) node[midway,left,text=TUMBlue] {\sm{790.3} \textcolor{TUMOrange}{($-$\sm{1.4})}};
		\draw[-latex] ([shift={(0.4,0)}]blocklist.south) -- ([shift={(0.4,0)}]gfw.north) node[midway, right,text=TUMBlue] {\sm{932.7} \textcolor{TUMOrange}{($-$\sm{1.5})}};

		\draw[-latex] ([shift={(-0.4,0)}]gfw.south) -- ([shift={(-0.4,0)}]apd.north) node[midway,left,text=TUMBlue] {\sm{656.1} \textcolor{TUMOrange}{($-$\sm{134.2})}};
		\draw[-latex] ([shift={(0.4,0)}]gfw.south) -- ([shift={(0.4,0)}]apd.north) node[midway, right,text=TUMBlue] {\sm{798.6} \textcolor{TUMOrange}{($-$\sm{134.1})}};
		
		\draw[-latex] ([shift={(-0.4,0)}]apd.south) -- ([shift={(-0.4,0)}]unresponsive.north) node[midway,left,text=TUMBlue] {\sm{405.3} \textcolor{TUMOrange}{($-$\sm{250.8})}};
		\draw[-latex] ([shift={(0.4,0)}]apd.south) -- ([shift={(0.4,0)}]zmap.north) node[midway, right,text=TUMBlue] {\sm{540.3} \textcolor{TUMOrange}{($-$\sm{258.3})}};

		\draw[-latex] ([shift={(-0.4,0)}]unresponsive.south) -- ([shift={(-0.4,0)}]zmap.north) node[midway,left,text=TUMBlue] {\sm{5.6} \textcolor{TUMOrange}{($-$\sm{399.7})}};

		\draw[-latex] ([shift={(-0.4,0)}]zmap.south) -- ([shift={(-0.4,0)}]responsive.north) node[midway,left,text=TUMBlue] {\sm{3.2}};
		\draw[-latex] ([shift={(0.4,0)}]zmap.south) -- ([shift={(0.4,0)}]responsive.north) node[midway,right,text=TUMBlue] {\sm{5.6}};
		
		\node[inputentity, above left= 25pt and -50pt of blocklist] (dots) {\ldots};
		\node[inputentity, above=1pt of dots] (DNS) {DNS Resolution};
		\node[inputentity, above=1pt of DNS] (traceroutes) {Traceroutes};
		\node[inputentity, above= 1pt of traceroutes] (ripeatlas) {RIPE Atlas};
		\node[innerentity,minimum width=4.1cm, rounded corners=6pt, above= 1pt of ripeatlas] (input) {Cumulative \hitlist};
		
		\draw[-latex] (dots) -- ([shift={(-0.4,0)}]blocklist) node [circle,inner sep=8pt,midway,left,text=TUMBlue] {\sm{791.7}};

		\node[inputentity, above right= 25pt and -50pt of blocklist,color=TUMDarkGreen] (tg) {\ldots};
		\node[inputentity, above=1pt of tg,color=TUMDarkGreen] (ps) {Passive Sources};
		\node[inputentity, above= 1pt of ps,color=TUMDarkGreen] (dc) {Target Generation};
		\node[inputentity, above=1pt of dc,color=TUMDarkGreen] (unresp_input) {30-Day Unresponsive};
		\node[innerentity, minimum width=4.1cm,rounded corners=6pt, above= 1pt of unresp_input,color=TUMDarkGreen] (newinput) {New Input Sources (\Cref{sec:new})};
		
		\draw[-latex] (tg) -- ([shift={(0.4,0)}]blocklist) node [circle,inner sep=8pt,midway,right,text=TUMBlue] {\sm{934.2}};

	\end{tikzpicture}%
	\caption{\hitlist pipeline. This work analyzes the existing steps and adds a \acl{gfw} filter (see \Cref{sec:hitlist}). Furthermore, the filtered, aliased prefixes are analyzed in \Cref{sec:aliased} and new input sources are evaluated in \Cref{sec:new}. Newly implemented service components are indicated by \textcolor{TUMDarkGreen}{green} borders.}
	\label{fig:hitlist_pipeline}
\end{figure}
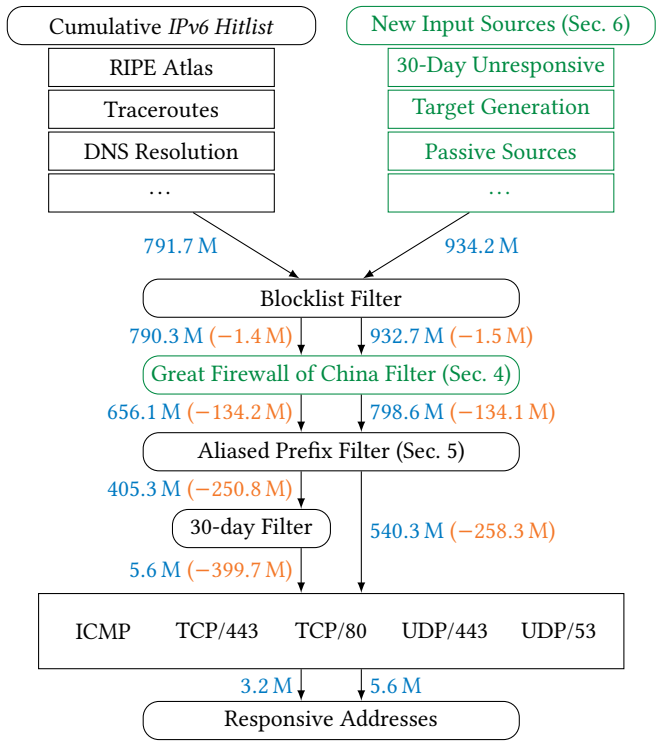

With the publication of the \hitlist service in 2018, Gasser \etal \cite{gasser2018clusters} established an ongoing service that collects IPv6 address candidates, identifies aliased prefixes, and tests the responsiveness in respect to different protocols, namely ICMP, TCP on port 80 (HTTP) and 443 (HTTPS), and UDP on port 53 (DNS) and 443 (QUIC) (see \Cref{fig:hitlist_pipeline}).
The service is well maintained, up to date and has been used as de-facto standard for IPv6 analysis and scans, \eg \cite{zirngibl2021over9000,aschenbrenner2021mptcp,nosyk2022loops,almeida2020loadbalancing,deccio2020closeddoors,rodday2021defaultroutes}.
However, it has not seen significant updates or analyses since its initial publication.
Changes in the usage of IPv6 and input sources might have influenced the quality of the service since 2018.
Furthermore, a variety of methodologies to \textit{generate} likely responsive addresses emerged \cite{liu20196Tree,cui20206GCVAE,yang20226Graph,cui20216GAN,cui20216VecLM} but have not been established as ongoing service publishing data.

\vspace{0.4em}
\noindent
\emph{Our main contributions in this work are:}

\first We evaluate the development of the \hitlist over the last four years and new biases introduced by the accumulation of new addresses. Our findings allow us to filter targets incorrectly tested as responsive. We identify \sm{134} addresses falsely reported as responsive to UDP/53  by the \hitlist since 2018 due to the \acl{gfw}'s DNS injection.

\second We analyze aliased prefixes in more detail and investigate whether the initial definition of a single host responsive to a complete prefix remains correct or whether a set of addresses needs to be treated differently.
We show that aliased prefixes host at least \sm{15} domains including ranked domains from different top lists \cite{alexa,majestic,umbrella}. In combination with additional findings, we suggest users of the hitlist to include subsets of these prefixes in future research.

\third We evaluate additional input sources and address generation mechanisms, and whether we can identify new, responsive addresses to improve the existing hitlist and support future research.
We identify \sm{5.6} new, responsive addresses, potentially improving coverage of the \hitlist by \sperc{174}.

\fourth We update the \hitlist service \cite{gasser2018clusters} to allow future research to use our findings within the established service.
\begin{center}
	\url{https://ipv6hitlist.github.io/}
\end{center}

\vspace{0.4em}
We present related work in \Cref{sec:related} and introduce used data sources in \Cref{sec:data}.
In \Cref{sec:hitlist}, we analyze the development of IPv6 and the \hitlist service \cite{gasser2018clusters} in detail.
We analyze aliased prefixes in \Cref{sec:aliased}, followed by an evaluation of potential new input sources in \Cref{sec:new}.
Finally, we discuss our findings in \Cref{sec:discussion} and conclude in \Cref{sec:conclusion}.

\Crefname{section}{Sec.}{Secs.}
\Crefname{section}{Section}{Sections}
\section{Related Work}
\label{sec:related}

Initial attempts to establish IPv6 hitlists have been conducted, \eg by \citeauthor{gasser2016scanning} \cite{gasser2016scanning} in 2016 and \citeauthor{fiebig2018somethingfromnothing} \cite{fiebig2018somethingfromnothing} in 2017.
\citeauthor{gasser2016scanning} \cite{gasser2016scanning} collected addresses from passive traces and active sources such as traceroutes and \ac{dns} resolutions, while \citeauthor{fiebig2018somethingfromnothing} \cite{fiebig2018somethingfromnothing} relied on reverse DNS to identify used addresses.
In 2016, \citeauthor{foremski2016entropyip} \cite{foremski2016entropyip} modeled IPv6 addresses based on their entropy and derived \emph{Entropy/IP}, relying on structural similarities between addresses to generate target candidates based on address seeds.
Based on these findings, \citeauthor{murdock20176Gen} \cite{murdock20176Gen} developed 6Gen in 2017 and derived an IPv6 hitlist containing \sm{55} active addresses.
However, they identify \sperc{98} of these as aliased.
In 2018, \citeauthor{gasser2018clusters}  \cite{gasser2018clusters} combined the earlier approaches and analyzed the composition of the resulting hitlist and its value.
They showed biases within different data sources and the importance of detecting aliased prefixes to remove additionally induced biases.
Based on this work, they established an ongoing service accumulating addresses from a variety of services and regularly testing their responsiveness.

Throughout the following years, different IPv6 address generation algorithms were published \cite{cui20206GCVAE,cui20216GAN,cui20216VecLM,hou20216Hit,liu20196Tree,yang20226Graph,song2022det}.
All approaches rely on the assumption that IPv6 addresses contain patterns due to assignment strategies, that allow research to guess new, responsive addresses.
They differ based on selected address representations and machine learning approaches.
\citeauthor{cui20216GAN} \cite{cui20216GAN} use General Adversary Networks while \citeauthor{liu20196Tree} \cite{liu20196Tree} represent addresses as a space tree.
The according studies mostly rely on the \hitlist \cite{gasser2018clusters} and evaluate their generated lists with active scans.
They were able to generate hitlists containing more than one billion candidates and reach hit rates of up to \sperc{50} \cite{song2022det}.
However, as of May 2022, we only found a snapshot of generated addresses from DET \cite{song2022det} and other sources could not be verified or reused for further studies.
We use 6Tree \cite{liu20196Tree}, 6Graph \cite{yang20226Graph}, 6GAN \cite{cui20216GAN}, and 6VecLM \cite{cui20216VecLM} to generate addresses as new input sources in this work and show the potential of some algorithms.
While we are not able to reproduce their published hit rates, we show that some algorithms can be used as valuable new sources in addition to the existing \hitlist.

Besides the collection or generation of new address candidates, the detection of aliased prefixes is important to understand and remove induced biases to IPv6 hitlists.
Different alias detection methodologies have been proposed to detect different addresses of the same router \cite{luckie2013speedtrap,beverly2013tbt,marder2020apple,vermeulen2020ratelimiting,padmanabhan2015uav6} and to detect IPv4 and IPv6 siblings \cite{scheitle2017clockskew}.
However, most of these exploit side channels not available on every target.
Therefore, \citeauthor{murdock20176Gen} \cite{murdock20176Gen} proposed an alias detection for IPv6 hitlists based on the responsiveness of random addresses within prefixes of size /96.
\citeauthor{gasser2018clusters} \cite{gasser2018clusters} extended this idea to a multi-level aliased prefix detection on different prefix lengths.
Furthermore, they combine TCP/80 and ICMP probes together with results from previous days to account for probe timeouts.
They verified the effectiveness of their approach to identify aliases with TCP fingerprints.

\citeauthor{song2022det} \cite{song2022det} suggested using the \ac{tbt} introduced by \citeauthor{beverly2013tbt} \cite{beverly2013tbt} to evaluate real aliases. They show that some prefixes identified as aliased by the \hitlist might contain multiple hosts, although being fully responsive.
In this work, we combine detected aliases based on the multi-level detection method, with TCP fingerprinting, the \ac{tbt}, and information about specific \acp{as} and hosted domains to shed further light on identified aliased prefixes.

While the \acl{gfw} does not necessarily seem related to IPv6 hitlists, we show its relation and impact in \Cref{sec:hitlist}, mainly its DNS injection behavior.
It has been analyzed in a different context \cite{farnan2016poisoningthewell, anonymous2014comprehensivefirewallpicture,anonymous2020tripletcensors} and has been recently used as a side channel to actively analyze DNS root server performance in China by \citeauthor{zhang2022DNSRootserver} \cite{zhang2022DNSRootserver}.
Most importantly, these works find that the \ac{gfw} is used to inject \ac{dns} responses for censored domains at the border of Chinese networks.
\citeauthor{anonymous2020tripletcensors} \cite{anonymous2020tripletcensors} analyzed the behavior in 2020 observing multiple responses that can be mapped to different injectors.
They further find that erroneous responses normally contain generally routed, valid IP addresses.
However, these addresses can be mapped to operators unrelated to the requested domain.
We identify the \ac{gfw} to be responsible for a majority of addresses responsive to DNS probes.
Results show similar behavior as reported by related work, highly impacting the quality of the \hitlist.
However, we currently observe different addresses in all responses.

\section{Data Sources and Scans}
\label{sec:data}

Our work relies on different public data sources but also individual scans.
We introduce all sources and used tools in the following section.
For our own scans, we strictly follow ethical considerations as explained in \Cref{sec:ethics}.

\subsection{IPv6 Hitlist Service}
Our primary source is the \hitlist published by Gasser \etal \cite{gasser2018clusters}.
The service collects address candidates from multiple sources, including DNS AAAA resolutions, conducted traceroutes, and public sources, \eg from RIPE Atlas and CT logs.
The service frequently \first updates addresses, \second uses \emph{all} collected addresses as input, \third applies multiple filters, and \fourth tests the responsiveness of addresses in respect to different protocols.
\Cref{fig:hitlist_pipeline} shows the service pipeline.

The first filter removes the addresses of operators who requested exclusion of regular scans following ethical considerations described in \Cref{sec:ethics}.
The most important existing filter is the aliased prefix detection.
The initial definition of aliased prefixes describes a single host responsive for all addresses in a prefix.
Each individual aliased prefix may be infeasible to scan, offers limited value to following scans, and introduces a bias.
As shown in \Cref{sec:aliased}, aliased prefixes in our data had different lengths between /28 and /120.
While earlier work from \citeauthor{murdock20176Gen} \cite{murdock20176Gen} tests for aliased prefixes with a fixed length of /96, the aliased prefix detection of the \hitlist tests prefixes of different lengths, including:
\begin{itemize}
	\item IPv6 prefixes announced in BGP;
	\item all /64 prefixes with at least one address contained in the \hitlist service input;
	\item and prefixes longer than /64 (in steps of four bit) with at least 100 addresses.
\end{itemize}

The implemented detection \cite{gasser2018clusters} relies on the assumption that it is highly unlikely that multiple randomly selected addresses within an IPv6 prefix are responsive.
Therefore, the detection selects one random address within each of the 16 more specific prefixes (0-f) and uses ZMapv6 \cite{zmapv6} to test responsiveness.
To test whether prefix \texttt{2001:db8::/32} is aliased, a single, random address is tested within all subprefixes \texttt{2001:db8:[0-f]000::/36}.
This address generation distributes the pseudo-random targets evenly across the complete prefix.
If all 16 addresses are responsive, the prefix is labeled as aliased.
ICMP and TCP/80 are tested, and results are merged across protocols and with the previous three scans. 
This reduces mis-classification of prefixes, \eg due to random network events or packet loss during individual scans.

The final filter in \Cref{fig:hitlist_pipeline} removes all addresses that are unresponsive for at least 30 days.
This filter reduces the required scan load drastically.
However, these addresses are never tested for responsiveness again after exclusion.
We re-scanned these addresses to test whether addresses are responsive after 30 days again as discussed in \Cref{sec:new}.

After all filter steps, the service executes traceroutes using \yarrp \cite{beverly2016yarrp} to all targets to potentially identify new targets.
Furthermore, ZMapv6 scans ICMP, TCP port 80 (HTTP) and 443 (HTTPS), and UDP port 53 (DNS) and 443 (QUIC).
While scans were executed on a daily basis initially, the growth of the input set increased the overall runtime to several days.
The data covers more than 750 scans between July 2018 and April 2022.

\subsection{Additional Data Sources}

Besides the \hitlist service, we use additional data throughout this work to analyze the development of the hitlist and improve its current state for future research.

\paragraph{\textbf{CAIDA Ark}}
While the  \hitlist contains its own traceroutes and data from RIPE Atlas, traceroutes conducted on a regular basis from the CAIDA Archipelago (Ark) Infrastructure are not yet included \cite{caidaArk}.
We use a snapshot of the data from March 2022 to analyze the value additional vantage points can offer to reveal new routers or whether the existing data set covers most contained targets.

\paragraph{\textbf{DET}}
\citeauthor{song2022det} \cite{song2022det} collected IPv6 addresses from different services similar to \citeauthor{gasser2018clusters} \cite{gasser2018clusters} and additionally used target generation algorithms.
They advertise an ongoing service and data publication.
However, we are only able to download a single snapshot of responsive addresses and did not receive a reply requesting additional data.
Thus, we only use this snapshot as a new source of addresses and evaluate its value.

\paragraph{\textbf{IPv6 Scans}}
Besides the publicly available data from the \hitlist service, we use the existing pipeline to conduct additional scans testing new IPv6 address sources and target generation methodologies analyzed in \Cref{sec:new} (see \Cref{fig:hitlist_pipeline}).
We use ZMapv6 with the same probe modules, configurations and payloads as the \hitlist service from the same vantage point.
We further deploy the multi-level aliased prefix detection from \citeauthor{gasser2018clusters} \cite{gasser2018clusters} and filter our scans with already known aliased prefixes and blocklists collected by the existing service.

\paragraph{\textbf{DNS Scans}}
We received access to DNS scans regularly conducted by our institution.
While these scans have not been explicitly set up for this study, ethical considerations described in \Cref{sec:ethics} are still applied.
These scans are already used as input for the existing \hitlist service.
The scans resolve domains from various input sources covering more than \sm{300} domains from the \ac{czds}, including \texttt{.com}, \texttt{.net}, and \texttt{.org}, Certificate Transparency logs, and cc-TLDs.
Additionally, the Alexa \topm \cite{alexa}, Majestic \topm \cite{majestic}, and Umbrella \topm \cite{umbrella} are resolved.

These scans resolve domains to IPv6 addresses (AAAA records)  but also to name server (NS) and mail exchanger (MX) records.
These resource records are resolved to their respective IPv6 addresses.
While the direct AAAA resolution of domains is already used as input of the \hitlist service, the name server and mail exchanger domains were not explicitly included.
We use a single snapshot of these scans from April 7th, 2022 to analyze aliased prefixes in \Cref{sec:aliased} and evaluate the quality of these domains as new input in \Cref{sec:new} and include them into the \hitlist service.

\subsection{Ethics}
\label{sec:ethics}

We primarily rely on available data from the \hitlist service.
However, some analyses are based on new scans, \eg the study of aliased prefixes and new IPv6 address sources.
During all scans, we follow ethical measures, \ie informed consent~\cite{menloreport} and community best
practices~\cite{PA16}. 
We scan with a limited rate and use a request-based blocklist.
Furthermore, we use the current blocklist of the existing \hitlist service to seed our blocklist.
This prevents interferences with networks that already requested to opt-out of existing IPv6 scans.
Our measurement vantage point is clearly identified.
This includes reverse \ac{dns}, WHOIS information and a hosted website informing about our measurements.
We did not receive any inquiries related to our scans during this work.

Additional \ac{dns} scans we rely on are conducted independently of this work but follow the same ethical principles, including a blocklist and public information about the scans and vantage point.

\section{IPv6 Hitlist Development}
\label{sec:hitlist}

As explained in \Cref{sec:data}, the \hitlist \cite{gasser2018clusters} is implemented as an ongoing service, regularly updating its input, identifying aliased prefixes (analyzed in more detail in \Cref{sec:aliased}), and testing the responsiveness of addresses.
We analyze the overall hitlist and responsiveness of addresses in the following.

\subsection{Input Development}
\label{sec:input_dev}

\begin{figure}
	\includegraphics{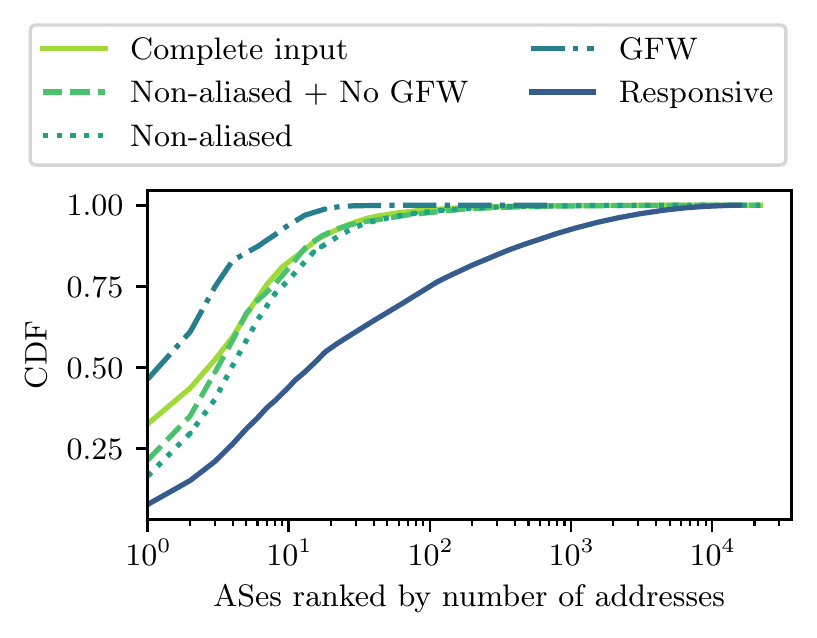}
	\caption{Distribution of addresses in the input list across ASes. Additionally, the effect of the alias and GFW filter are displayed. Note the log x-Axis. While the complete input is biased towards some ASes, established and new filters result in a well distributed set of responsive addresses.}
	\label{fig:complete_asn_cdf}
\end{figure}

Starting with \sm{90} addresses in July 2018, the service has accumulated more than \sm{790} addresses until April 2022.
They cover \num{22074} \acp{as} compared to \num{10866} in 2018 \cite{gasser2018clusters}.
The \sk{22} \acp{as} cover \sperc{76} of \acp{as} announcing at least one IPv6 prefix in BGP at the time of writing.
The number of announced prefixes and \acp{as} is based on a routing information base from RIPE RIS collector rrc00 \cite{riperis}.\footnote{\url{https://data.ris.ripe.net/rrc00/2022.04/bview.20220407.0800.gz}}
Furthermore, the current input list covers four times more (\sk{97}) announced BGP prefixes compared to 2018, \sperc{62} of all announced prefixes.
The visible growth is similar to the general growth of IPv6 deployments and usage on the Internet and shows that the hitlist is able to adapt to it.

\Cref{fig:complete_asn_cdf} shows the cumulative distribution of IPv6 addresses in the hitlist input across \acp{as}.
Without any filter, the most prominent \ac{as} is Amazon (AS16509), covering \sperc{32} of all addresses.
However, \sperc{99.6} of these are filtered due to the aliased prefix detection.
\Cref{sec:resp} explains the additional plots covering the \ac{gfw} impact and responsive addresses.

Nevertheless, after filtering aliased prefixes, \sperc{80} of the current input is still covered by only 10 \acp{as}, mostly from \acp{isp} such as ANTEL (AS6057, \sperc{16}) or DTAG (AS3320, \sperc{10}).
Analyzing the source of these addresses reveals that they are mostly accumulated due to regular traceroutes, especially from RIPE Atlas, changing prefix assignments and rotating addresses.
\sm{282} addresses from the input contain a EUI-64 \ac{iid} including \texttt{ff:fe} and are based on a MAC address.
Extracting the \ac{iid} reveals that these addresses are only derived from \sm{22.7} distinct MAC addresses.
Grouping addresses based on the EUI-64 values shows that \sm{9} occur only within one IPv6 address each, while the remaining are seen in multiple addresses.
The most frequent EUI-64 value can be seen in \sk{240} distinct IPv6 addresses.
The \ac{oui} of the MAC address is mapped to a vendor (ZTE) and all addresses are part of the same /32 prefix, but within different subnets.
Similarly, \citeauthor{rye2021followthescent} \cite{rye2021followthescent} found a variety of \ac{cpe} used EUI-64 \acp{iid} and whose ISPs regularly rotated prefixes.
They were able to track individual devices within \acp{isp}.

These rotating addresses result in a visible bias of the complete input list towards a small set of \acp{as} due to the ongoing accumulation of addresses.
While the filter of unresponsive addresses after 30 days removes most of these addresses from regular scans and the responsive results afterward, the overall input list has to be treated carefully by other research relying on this  data.

\subsection{Address Responsiveness}
\label{sec:resp}
\begin{figure*}
	\includegraphics[width=\linewidth]{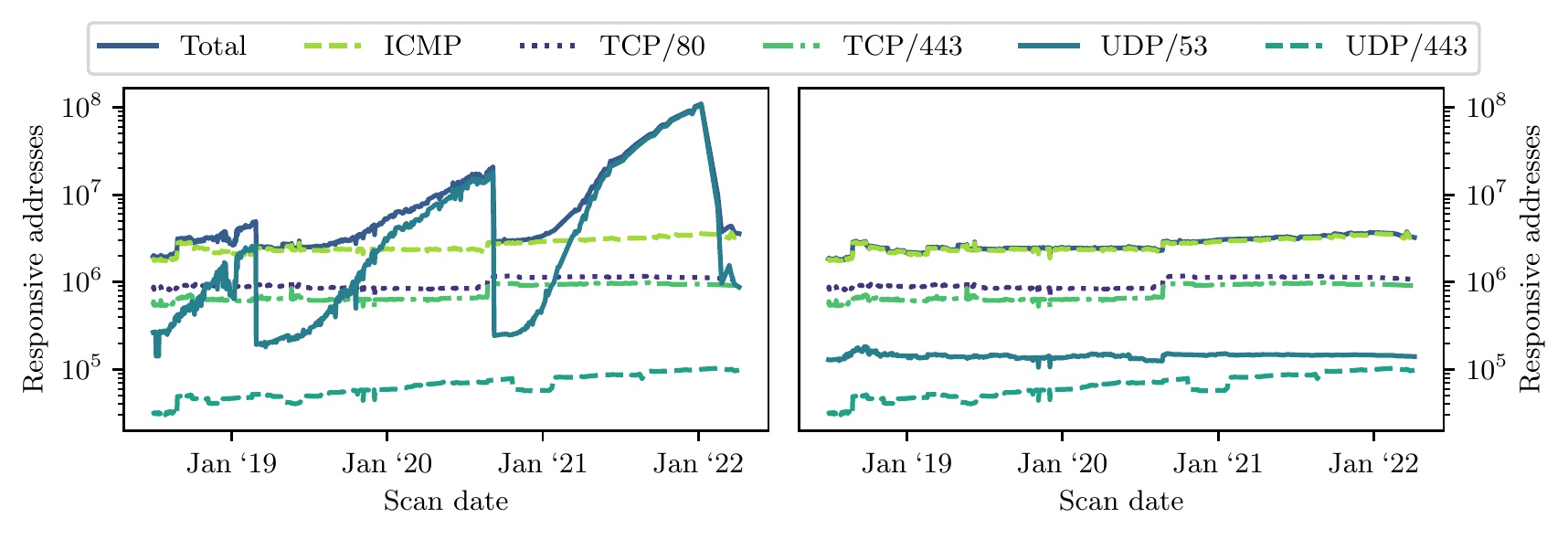}
	\caption{Comparison of the \hitlist as published (left) and cleaned from GFW injection (right). Without injected responses by the GFW, a steady development of the \hitlist and each protocol is visible. Note the log y-Axis.}
	\label{fig:hitlist}
\end{figure*}

\Cref{fig:hitlist} shows the responsiveness of addresses throughout the \hitlist service lifetime.
It depicts the responsiveness for each tested protocol and a total count of addresses responsive to at least one protocol within each scan. 
The left half represents the published state, with clearly visible spikes which will be explained in the following, and on the right, a \emph{cleaned} version is presented.

\paragraph{\textbf{The Prevalence of DNS}}
The published hitlist contains significant spikes in responsive addresses as seen in \Cref{fig:hitlist}.
During these events, the number of addresses responsive to \ac{dns} probes is larger compared to all other protocols.
After each peak, the number of addresses drops to a similar level as before the event.
The last event started in early 2021 with the most drastic growth of addresses, peaking at more than \sm{100}.
In comparison, only \sm{3.5} and \sm{1.4} addresses were responsive to ICMP and TCP/80 respectively, and no increase was visible during the same periods.
The peak drops in February 2022, after we implemented a filter based on the following findings and updated the service.

To understand the origin of these peaks, we describe the scan configuration and analyze results in detail.
The service sends a DNS query requesting a AAAA record for \texttt{www.google.com}.
All peak events share similarities but differ slightly.
During the first two events, a significant fraction of addresses responded with A records only containing an IPv4 address.
During the third and most recent event, responses carried valid AAAA records but contained Teredo addresses.
Note that Teredo is a deprecated standard embedding an IPv4 into an IPv6 address \cite{rfc4380}.
Furthermore, \zmap accumulated two or three responses for each scanned address, with up to 440 responses in the worst case.

Analyzing the erroneous IPv4 addresses contained in A records during earlier events and embedded in the returned Teredo addresses in the latter event reveals that none can be associated to the Google \acp{as} but other companies like Facebook, Microsoft or Dropbox.
Collecting all IPv6 addresses that responded with a clearly erroneous record (IPv4 or Teredo address) throughout the four years accumulates to more than \sm{134} addresses (\sperc{17} of the cumulative \hitlist input on April 7th, 2022).

Querying a different domain shows that these targets are not responsive themselves, but responses are injected and falsely interpreted as success by \zmap.
Most addresses with this response behavior are announced by \acp{as} of Chinese networks, \eg China Telecom Backbone (AS4134) and China Telecom (AS4812) originating \sperc{46.44} and \sperc{14.59} of impacted addresses respectively. 
\Cref{fig:complete_asn_cdf} compares the AS distribution of IPv6 addresses responding with these incorrectly responded addresses to the complete input.
These addresses cover only 695 ASes.
\sperc{93} of addresses are located in only 10 Chinese \acp{as}.
See \Cref{tab:appendix_china_as} in \Cref{app:as} for an overview of the Top 10.
We used MaxMind GeoLite2 \cite{maxmind} as an additional indicator of network location.
While we are aware of potential inaccuracies especially on a city level \cite{poese2011geolocation,scheitle2017hloc}, it mapped a majority of impacted IPv6 addresses to China.
Given these indicators a strong relation of these addresses to Chinese networks can be seen.

The overall behavior has been described similarly in related work \cite{anonymous2020tripletcensors,anonymous2014comprehensivefirewallpicture,farnan2016poisoningthewell}.
We see multiple responses to a single query indicating multiple injectors, responses are mostly in relation to addresses from China and \texttt{www.google.com} is a blocked domain.
Querying different blocked domains from these addresses shows similar behavior.
In contrast, a domain owned by ourselves, most likely not blocked, results in no response at all, not even a \ac{dns} error.
A difference to existing related work is that the currently injected responses carry a Teredo address not explicitly reported in previous findings \cite{farnan2016poisoningthewell, anonymous2014comprehensivefirewallpicture,anonymous2020tripletcensors}.
However, contained IPv4 addresses show similar behavior (cf.\cite{anonymous2020tripletcensors}) and can be mapped to previously identified incorrect networks.

The source of these addresses is the regularly conducted traceroutes by the \hitlist service using \yarrp.
Traceroute captures regularly changing addresses mostly with randomized \acp{iid} and visible as the last responsive hop.
The targeted address is not responsive itself.
Scanning these addresses in the following with \zmap triggers a \ac{dns} injection by the \ac{gfw}, but for a majority no other protocol is responsive.
In some cases, these targets are actually responsive for other probed protocols.
Thus, invalid \ac{dns} responses should be filtered but individual addresses should remain in the \hitlist if responsive to other protocols.

\begin{table*}
	\centering
	\caption{Development of responsive IPv6 addresses and covered \acp{as} over four years. Results are based on cleaned data, removing \ac{gfw} injected responses. For each year, a snapshot representing a single scan is used. The cumulative result covers all scans since the start of the \hitlist.}
	\label{tab:ip_as}
	\begin{tabular}{lrrrrrrrrrrrr}
		\toprule
		Year &     \multicolumn{2}{c}{ICMP} &  \multicolumn{2}{c}{TCP/443} &    \multicolumn{2}{c}{TCP/80} &  \multicolumn{2}{c}{UDP/443} &   \multicolumn{2}{c}{UDP/53} & \multicolumn{2}{c}{Total} \\
		\cmidrule(lr){2-3}\cmidrule(lr){4-5}\cmidrule(lr){6-7}\cmidrule(lr){8-9}\cmidrule(lr){10-11} \cmidrule{12-13} 
		& Addr. & ASes & Addr. & ASes & Addr. & ASes & Addr. & ASes & Addr. & ASes & Addr. & ASes \\
		\midrule
		2018-07-01 &  \sm{1.7} & \sk{10.1} &  \sk{550.6} &    \sk{5.8} &   \sk{832.1} & \sk{6.2} &   \sk{31.0} & \sk{0.9} &  \sk{129.1} & \sk{5.1} & \sm{1.8} & \sk{10.3} \\
		2019-04-01 &  \sm{2.4} & \sk{11.0} &  \sk{645.8} &    \sk{6.2} &   \sk{919.2} & \sk{6.6} &   \sk{50.4} & \sk{1.0} &  \sk{145.4} & \sk{5.2} & \sm{2.5} & \sk{11.2} \\
		2020-04-01 &  \sm{2.3} & \sk{11.7} &  \sk{632.8} &    \sk{6.6} &   \sk{836.2} & \sk{6.9} &   \sk{67.7} & \sk{1.3} &  \sk{148.4} & \sk{5.1} & \sm{2.4} & \sk{11.9} \\
		2021-04-02 &  \sm{3.0} & \sk{13.7} &  \sk{954.8} &    \sk{7.4} &   \sm{1.1}   & \sk{7.7} &   \sk{83.0} & \sk{1.3} &  \sk{148.0} & \sk{6.0} & \sm{3.1} & \sk{13.9} \\
		2022-04-07 &  \sm{3.1} & \sk{15.4} &  \sk{910.8} &    \sk{7.9} &   \sm{1.0}   & \sk{8.2} &   \sk{98.1} & \sk{2.0} &  \sk{140.7} & \sk{6.0} & \sm{3.2} & \sk{15.7} \\
		\midrule
		Cumulative & \sm{45.3} & & \sm{6.7} & & \sm{8.6} &  & \sm{2.5} & & \sk{200} & & \sm{46.8} &  \\
		\bottomrule
	\end{tabular}
\end{table*}

\paragraph{\textbf{Reducing the \acl{gfw} Impact}}
The \hitlist service sends DNS queries for a blocked domain, namely \texttt{www.google.com}.
These queries result in \ac{dns} injections from the \ac{gfw} which were labeled as successful responses by \zmap.
While changing the queried domain could prevent this, \eg to \texttt{example.com}, we decided to operate the \hitlist service with its current configuration.
This consistency in the service behavior increases the comparability of results over time.
With a consistent scan, different results indicate changes in the target behavior and are not induced by the scan itself.

This decision to focus on consistency requires an ongoing data cleaning step.
We initially implemented a filter removing \sm{134} IPv6 addresses to reduce the \ac{gfw} impact.
The \hitlist service saw at least one DNS injection for these addresses during July 2018 and April 2022 but no response to any other protocol.
This filter immediately reduced scan duration and impact on the Internet.

To reduce the future impact on the ongoing \hitlist service, we filter the DNS/53 results directly after the scan. 
Therefore, the results correctly reflect the responsiveness of newly scanned input addresses.
Thus, if addresses are responsive to any other protocol, they remain in the scan input. 
Otherwise, they are filtered by the 30-day filter (see \Cref{fig:hitlist_pipeline}).

\paragraph{\textbf{Evaluation of Remaining IPv6 Addresses Supporting DNS}}
After removing \ac{gfw} injected \ac{dns} responses, \sk{140} addresses responsive to \ac{dns} probes remain.
We evaluate the quality of the remaining IPv6 addresses with an experiment including a domain under our control.
To analyze the behavior of each scanned target, we query for a subdomain including a unique hash instead of a static domain.
The name server returns a valid AAAA record for the requested domain.
This approach allows us to map outgoing probes towards individual addresses to incoming requests at our name server.
Within this filtered set of targets, \sk{131.8} (\sperc{93.8}) of probes result in valid DNS responses with status codes indicating errors, because these targets are either name servers or resolvers unwilling to resolve the requested domain recursively.
\sk{6.5} (\sperc{4.6}) return regular responses with the correct AAAA record and according requests from the same IPv6 addresses are visible at our name server.
593 targets respond with a referral to name servers of the root or  our domain's parent zone.
Only 15 IPv6 addresses return a correct record, but the source addresses of incoming requests on our name server do not match the probed targets, \eg due to proxies or the usage of another interface.

The remaining \sperc{1.1}  of targets respond incorrectly but an analysis of results reveals nothing similar to the \ac{gfw} injection.
Responses contain for example incorrect status codes or referrals to \textit{localhost}.

\paragraph{\textbf{Cleaned \hitlist}}
We cleaned the historical data from \ac{gfw} injected responses resulting in the timeline shown in the right part of \Cref{fig:hitlist}.
This results in a relatively stable number of responsive addresses for all protocols respectively, each with a slight increase throughout the four-year period.
This development is in line with statistics reported by RIPE NCC regarding the general growth of IPv6 assignments \cite{ripev6growth}.
Most addresses responsive to at least one protocol are responsive to ICMP, followed by TCP/80 and TCP/443.
\Cref{fig:resp_overlap} in \Cref{app:responsive} shows the exact overlap between protocols.
For ICMP, the responsiveness increased from \sm{1.78} addresses covering \sk{10.1} \acp{as} to \sm{3.15} addresses in \sk{15.5} \acp{as}.
Results for the remaining protocols can be seen in \Cref{tab:ip_as}.

While UDP/443 (QUIC) increased the most by a factor of three, it still shows the worst response rate below UDP/53, even though QUIC was finally standardized in May 2021 \cite{rfc9000}.
Interestingly, \citeauthor{zirngibl2021over9000} \cite{zirngibl2021over9000} reported \sk{210} IPv6 addresses supporting QUIC in 2021, more than twice as many addresses as reported by the \hitlist.
However, they include addresses from DNS resolutions many of which are located in aliased prefixes (\eg from Cloudflare, Fastly, and Amazon) and thus filtered by the hitlist service.
We analyze this effect in more detail in \Cref{sec:aliased}.

While an overall growth is visible throughout the four years, the number of responsive IPv6 addresses to ICMP, TCP/80 and TCP/443 slightly decreases in between the analyzed scans from 2019 and 2020 (see \Cref{tab:ip_as}).
This is primarily due to input sources only added once, \eg rDNS data.
A fraction of these addresses are not responsive anymore after several scans.
However, without updates of the input, newly responsive addresses are not discovered.

\Cref{fig:complete_asn_cdf} shows the distribution of responsive addresses across \acp{as}.
Compared to the overall input, the distribution is flatter. 
The top \ac{as}, Linode (AS63949), covers only \sperc{7.9}, followed by China Telecom (AS4812).
\sperc{50} of responsive addresses are covered by 14 \acp{as}.
A distribution of addresses responsive to individual protocols can be seen in \Cref{fig:asn_resp_proto} in \Cref{app:responsive}.

Whereas the set of responsive addresses changes over time, the cardinality of the set remains relatively stable.
Since July 2018, \sm{46.8} non-aliased addresses  have been responsive at least once and to one protocol.
This is primarily due to ICMP with \sm{45.3} (\sperc{96.8}) addresses followed by TCP/80 with \sm{8.6} (\sperc{18.4}).
This shows that frequent change is visible, and an up-to-date service is necessary to provide a high-quality service for further research.

\begin{figure}
	\includegraphics[width=\linewidth]{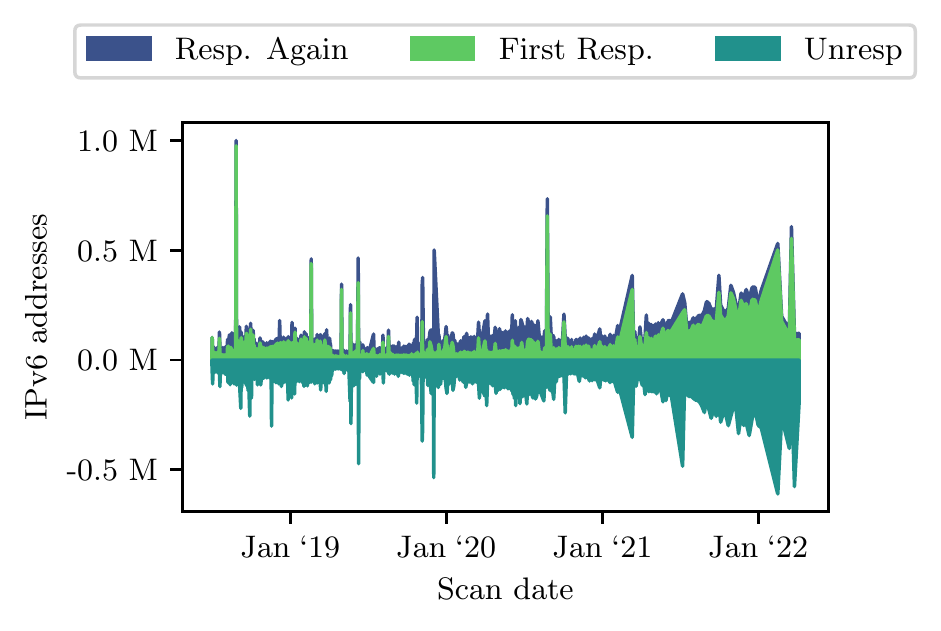}
	\caption{Development of responsive addresses over time. Later scans have a larger runtime covering up to 7 days, increasing the seen churn in between scans.}
	\label{fig:response_history}
\end{figure}

\Cref{fig:response_history} indicates the stability of the responsive address set over time.
It shows how many addresses are newly responsive or unresponsive for each scan compared to the previous scan.
In case of new IPv6 addresses, the figure differentiates between addresses that are completely new or were already responsive in a previous scan but not the last.
Generally, a frequent churn of \sk{200} to \sk{500} IPv6 addresses can be seen between two consecutive scans within one to five days.
However, unresponsive addresses are frequently recurring afterward.
Nevertheless, completely new IPv6 addresses are regularly visible.
As explained in \Cref{sec:data} the \hitlist service is not executed daily anymore but takes multiple days resulting in the increased churn in between scans towards the end of the analyzed period as seen in \Cref{fig:response_history}.

More significant changes are either due to issues with the aliased prefix detection, if a drop is directly visible after a large increase or due to missing sources for a scan if the drop is before the rise.
Increases without a close drop are primarily due to added sources, \eg an addition of IPv6 addresses from rDNS scans as used by \citeauthor{fiebig2018somethingfromnothing} \cite{fiebig2018somethingfromnothing}.

\subsection{\textit{Takeaways and Suggestions}}

The analysis of the current hitlist state shows that a regularly updated, ongoing service is required to provide an up-to-date view and adapt to regular changes within the Internet.
While \sk{176.6} addresses were responsive throughout the entire period (\sperc{5.4} out of \sm{3.2} on April 7th, 2022), regular churn is visible across all protocols.
On the other side, the ongoing accumulation of addresses leads to an overall input list containing unresponsive addresses.
Especially regular traceroutes identify a large set of addresses derived from a fraction of EUI-64 \acp{iid} but located within regularly changing prefixes (see \Cref{sec:input_dev}).

Furthermore, ongoing services require regular monitoring to understand the impact of network changes on the scanning methodology, \eg the impact of the \ac{gfw} that slowly ramped up over time and changed behavior frequently.
The \ac{gfw} injections additionally led to increased problems in the comparability of results.
Chinese vantage points are most likely affected by the \ac{gfw} injection as well but on the complete opposite set of addresses, namely targets outside Chinese networks.

\textit{We suggest frequent monitoring of the hitlist in the future by the operators but also the community.
Furthermore, we plan to frequently clean the overall input of specific addresses, such as outdated EUI-64 based addresses to better support scans of other protocols.}

\section{Aliased Prefix Analysis}
\label{sec:aliased}

Besides the difficulty of IPv6 scans due to the inherent problem of the large address space itself, challenges occur due to the fact that individual addresses do not necessarily identify individual targets.
With IPv6, servers can be reached using multiple IP addresses or even complete prefixes.
They often appear as fully used prefixes with each address responsive, \eg to ICMP or TCP handshakes.
The most commonly assumed reason is aliasing, where a \emph{single} target is reachable using a complete prefix.
The \hitlist tries to identify these prefixes as described in \Cref{sec:data}.
Regarding scans, this is mainly a problem with IPv6.
However, similar occurrences have also been mentioned in combination with IPv4 address scans, \eg by \citeauthor{izhikevich2021lzr} \cite{izhikevich2021lzr} or \citeauthor{alt2014v4aliases} \cite{alt2014v4aliases}.

The implemented mechanism by the \hitlist tests for responsiveness.
However, identified prefixes do not have to be used by a single endpoint, \ie as an alias.
Other reasons can be for example \acp{cdn}, using complete IP prefixes for multiple servers or middle-boxes and proxies preemptively terminating connection attempts.
Especially in the case of \acp{cdn}, (a subset of) these fully responsive prefixes might be a valuable input for advanced scans, \eg to analyze protocol deployments such as QUIC or TLS1.3.
In the following, we investigate these address regions in more detail to allow a better understanding of the \hitlist and a better usage of its outcome.

In 2018, the \hitlist service identified \sk{12} aliased prefixes of different sizes.
The number of aliased prefixes increased steadily throughout the year, reaching \sk{42.8} in January 2022 and a sudden increase to \sk{111.5} prefixes afterward.
The latter growth by \sk{66.4} prefixes is due to a single \ac{as}, namely Trafficforce (AS212144), a Lithuanian network starting to announce a larger number of prefixes in February 2022, solely limited to IPv6.
All aliased prefixes were /64, responding to ICMP but not TCP/80.
This sudden increase by a single operator stands out.
The classification was based on successful ICMP probes and is reproducible.
We contacted the technical support of the \ac{as} regarding this steep increase but did not receive further information.

\begin{figure}
	\includegraphics[width=\linewidth]{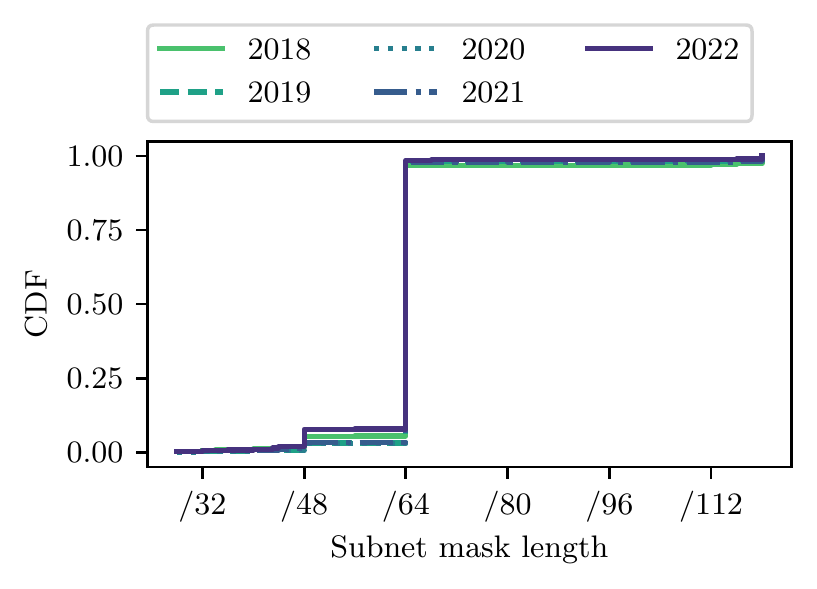}
	\caption{Distribution of aliased prefix sizes over time. The plot for 2022 excludes Trafficforce (AS212144) accounting for 60~k (61.6~\%) aliased prefixes mostly /64. For each year, a snapshot representing a single scan is used. More than 90~\% of aliased prefixes had a length of /64.}
	\label{fig:aliased_sizes}
\end{figure}

We analyze whether the increasing number of addresses in the overall hitlist (cf. \Cref{sec:hitlist}) results in the remaining increase from \sk{12} to \sk{42.8} labeled prefixes.
As explained in \Cref{sec:data}, the service only checks for aliased prefixes of size /68 and longer, if more than 100 addresses from this prefix are part of the \hitlist service input.
Thus, aliased prefixes might remain unrecognized if too few addresses were in the input.
Therefore, we check the size of aliased prefixes shown in \Cref{fig:aliased_sizes}.
It shows the distribution of aliased prefixes across prefix sizes for a single snapshot each year.
It has been similar throughout the years with a small percentage of prefixes within /28 and /60.
The shortest aliased prefixes are several /28s announced by EpicUp (AS397165) a US-based cloud provider.
However, most aliased prefixes constantly have a length of /64 where the hitlist does not require a threshold.
For this prefix length, only a single address is required to trigger the detection mechanism.
Newly identified prefixes throughout the years were either not aliased in 2018 or did not contain a single address.
Therefore, testing /64 prefixes if a single address is known to the \hitlist proves to be effective to detect a majority of aliased prefixes.

\subsection{Fingerprinting Aliased Prefixes}
We use fingerprinting approaches to analyze aliased prefixes identified by the \hitlist service in more detail.
These methodologies allowed us to evaluate whether aliased prefixes were a single responsive host, or whether some identified aliased prefixes contained multiple hosts.
We combine two methods to fingerprint hosts in identified aliased prefixes, namely fingerprinting based on TCP response features as conducted by \citeauthor{gasser2018clusters} \cite{gasser2018clusters} and the \acf{tbt} as presented by \citeauthor{beverly2013tbt} \cite{beverly2013tbt}.

\paragraph{\textbf{TCP Fingerprints}}
We use TCP fingerprints extracted from the scans for the aliased prefix detection \cite{gasser2018clusters}.
It relies on different features from the TCP handshake, namely the \emph{Optionstext}, an order preserving string representation of TCP options, the TCP window size and window scale option, the \ac{mss} and the iTTL.
The iTTL rounds the TTL to the next power of 2 \cite{mukaddam2014ittl,backes2016ittl} and thus represents the likely selected initial TTL.
Therefore, it reduces inconsistencies based on routing changes or middleboxes.
We omit the timestamp analysis as Linux machines using kernel 4.10 or newer randomize the value and are expected to show increased deployment since 2018.
In contrast to the \ac{tbt}, the same values between two IPv6 addresses do not necessarily indicate the same host, but similarly, varying values suggest different hosts.

TCP fingerprints can be derived for \sk{33.5} aliased prefixes.
The remaining prefixes were detected based on ICMP scans.
For \sk{33.3} (\sperc{99.5}) of these prefixes, all values match while for \num{160}, differences are visible.
The iTTL, \ac{mss}, window scale option and \emph{Optionstext} only differ for addresses within up to 13 prefixes.
In contrast, the TCP Window Size is different for addresses in 154 aliased  prefixes.
However, the Window size can change on a single host within different connections and different values are not necessarily a discriminating factor.
In comparison to the evaluation by \citeauthor{gasser2018clusters} \cite{gasser2018clusters} we see similar results.
Most fingerprintable prefixes respond with a uniform behavior, while a small fraction shows variable behavior.
This supports the initial assumption that detected fully responsive prefixes are often aliases for the same host.
However, some fingerprints indicate that some detected address blocks are potentially used differently and by multiple hosts.

\paragraph{\textbf{\acl{tbt}}}
We used the \ac{tbt} as an additional indication of prefix usage.
It relies on the characteristic of IPv6 that only end hosts are allowed to fragment packets and relies on a shared \ac{pmtu} between aliased addresses for the same host.
In general, if a router receives an IPv6 packet that is too big, it has to notify the sender using an ICMPv6 Packet Too Big Messages.
Afterward, the sender should update its \ac{pmtu} cache and fragment the respective packet.
While this process was initially utilized to identify alias addresses of routers, \citeauthor{song2022det} \cite{song2022det} proposed it to analyze aliased prefixes.
We shortly describe the required steps in the following:

\noindent
\first The \acl{tbt} verifies that a set of addresses (8) under test within a prefix replies to ICMP Echo Requests of size \SI{1300}{\byte}, slightly larger than the minimum required MTU for IPv6 of \SI{1280}{\byte}, without fragmentation.

\noindent
\second It sends ICMPv6 Packet Too Big Messages itself to \emph{one} of the addresses and verifies that the next round of ICMP Echo Request messages is in fact fragmented.

\noindent
\third It sends ICMP Echo Requests to the remaining addresses under test \emph{without} the preceding error message. In case all addresses are aliases for the same device and interface, they share the same \ac{pmtu} cache and should now fragment the response.

The methodology only provides insights if targets respond to the initial ICMP Echo Requests without fragmentation.
We used this methodology on the \sk{111} prefixes identified by the \hitlist service on April 7, 2022 and received successful results for \sk{29.4}.
Out of these, for  \sk{27.6} (\sperc{93.75}) all eight responses were fragmented after the initial error message to a single address indicating a shared \ac{pmtu} cache.
Only for 249 (\sperc{0.85}) prefixes, no request resulted in a fragmented response but each individual address required a new error message.

Interestingly within the remaining prefixes (1592, \sperc{5.4}), between two and seven addresses share a \ac{pmtu} cache but not all.
This effect is mostly seen with Akamai and Cloudflare with \sk{1} and 268 prefixes respectively.
This supports our assumption that a fraction of identified \textit{aliased} prefixes is not completely in line with the initial definition of a single host with the complete prefix as an alias.
Nevertheless, addresses from these fully responsive prefixes are still not assigned to single hosts used within a load balancing setup, \eg of \acp{cdn}.

\subsection{Characteristics of Aliased Prefixes}
Besides these technical fingerprints, we analyze additional characteristics of aliased prefixes in addition to the originating \acp{as}.
We argue that these characteristics allow for a more informed evaluation and usage of the \hitlist results in the future.

\begin{figure}
	\includegraphics[width=\linewidth]{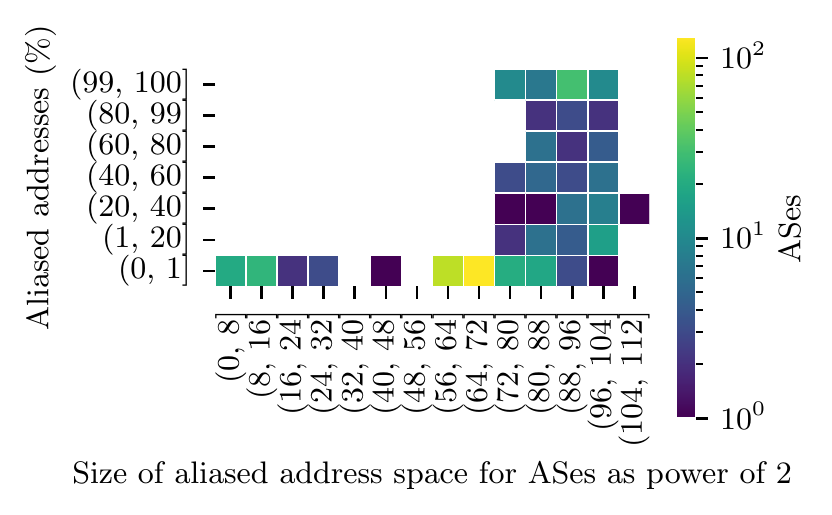}
	\caption{Number of aliased addresses (power of two) within ASes in relation to their overall number of announced addresses. If multiple prefixes within an AS are aliased, the number of addresses are summed. For some ASes, more than 99~\% of the announced IPv6 addresses were covered by aliased prefixes. Results are based on the aliased prefix detection of April 7th, 2022.}
	\label{fig:as_aliased_fraction}	
\end{figure}

\paragraph{\textbf{Are Individual Prefixes Fully Responsive or Complete ASes?}}
To analyze whether aliased prefixes are more likely due to individual network entities or set up on an \ac{as}-level, we analyze the fraction of aliased addresses within each \ac{as} in respect to the total number of announced IPv6 addresses by an \ac{as}.
\Cref {fig:as_aliased_fraction} shows this relation for all \acp{as} with at least one aliased prefix.
The total number of addresses from aliased prefixes is given as power of two on the x-axis ranging from $2^8$ to $2^{112}$ addresses.
This value does not necessarily represent single prefixes but sums all aliased prefixes within each \ac{as}.
The highest number of IPv6 addresses from aliased prefixes is again due to EpicUp (AS397165) announcing 61 fully responsive /28 prefixes.
The y-axis depicts the fraction in respect to all announced IPv6 addresses by the respective \ac{as}.
The axes are binned for better visibility.

While the fraction for many \acp{as} is less than 1\textperthousand, for 80 \acp{as} more than \sperc{50} of announced addresses are located in aliased prefixes, and for 61 \acp{as} even  more than \sperc{90} are reached.
The most prominent candidates in the latter category are Fastly (AS54113) with \sperc{95.3}, but also AS33905 owned by Akamai and AS209242 owned by Cloudflare both aliased to \sperc{100}.
We argue that in these cases even without exact fingerprinting, it is highly unlikely that all addresses are an alias of a single host because these \acp{cdn} serve numerous websites and clients.
The complete exclusion of all addresses might almost exclude complete \acp{as}.

\paragraph{\textbf{Are Domains Hosted in Aliased Prefixes?}}
To further evaluate the effect of an exclusion of all aliased prefixes, we analyze whether domains are hosted within these networks.
In the case of higher layer protocol evaluation, including TLS, QUIC or HTTP but also the analysis of Internet consolidation, these targets can be highly relevant and should not be excluded completely.
We use a single snapshot of our DNS scans introduced in \Cref{sec:data} from April 7th, 2022 to analyze whether identified, aliased prefixes host domains and to which extent.

Based on our data \sm{15.0} domains resolved to \sk{5.2} aliased prefixes in total.
133 different ASes announced these prefixes.
Note that this is a lower bound, given that we resolve a subset of the DNS namespace and that load balancing might impact our scans.
Nevertheless, it shows that a fraction of hosting infrastructure can be missed by research based on the \hitlist.
The most prominent \ac{as} with aliased prefixes hosting domains was Cloudflare (AS13335) with 115 prefixes each hosting a mean of \sk{167.0} domains.
The highest number of domains even reached \sm{3.94} domains within a single fully responsive /48.
\acp{cdn} such as Fastly (AS54113), Amazon (AS16509), and Google (AS15169) were affected.

We investigated how many of these domains were on top lists.
As explained in \Cref{sec:data}, three top lists were resolved, all containing \sm{1} domains respectively on April 7th, 2022.
Domains from all top lists resolved to IPv6 addresses within aliased prefixes:
\begin{itemize}
	\item Alexa \topm: \sk{177.0}
	\item Majestic \topm: \sk{170.2}
	\item Umbrella \topm: \sk{118.0}
\end{itemize}
Considering the Alexa \topm, 129 and \sk{22.6} domains were within the \topk and Top\,\sk{100} respectively, including \texttt{facebook.com} and \texttt{spotify.com}.
Affected domains from the Majestic \topm were of similar ranks.
In contrast, only 53 affected domains from the Umbrella top list were within the \topk.

Including addresses out of aliased prefixes contained in AAAA records would include \sm{2.7} distinct addresses, a small fraction of the respective aliased prefixes.
We argue that all of these or at least a subset of IPv6 addresses should be considered by researchers in the analysis of protocols on top of IPv6 even though they are identified as aliased prefixes.

\begin{table}
	\centering
	\caption{Responsiveness of aliased prefixes. For each prefix one random address is tested to reduce impact.}
	\label{tab:aliased_proto_scan}
	\begin{tabular}{lrr}
		\toprule
		Protocol & {\# Prefixes} & \# ASes \\
		\midrule
		ICMP	& \sk{39.0} & 270 \\
		TCP/443	& \sk{31.9}	& 155 \\
		TCP/80	& \sk{32.3}	& 179 \\
		UDP/443	& \sk{28.8} & 41 \\
		UDP/53	& 172 & 32 \\
		\bottomrule
	\end{tabular}
\end{table}

\begin{table*}
	\centering
	\caption{New input sources for IPv6 address candidates evaluated in this work. The covered \acp{as} are set in relation (\%) to the total number of \acp{as} (29~k) announcing IPv6 prefixes based on RIPE RIS \cite{riperis}.}
	\label{tab:new_sources}
	\begin{tabular}{llrrr}
		\toprule
		& & & \multicolumn{2}{c}{ASes} \\
		\cmidrule{4-5}
		Source & Information & Addresses & Total  & \%\\
		\midrule
		Passive sources & Extracted addresses from e.g., MX/NS records, CAIDA Ark \cite{caidaArk}, DET \cite{song2022det} & \sk{356.7} & \sk{3.6} & 12.5  \\
		Unresponsive addresses & All addresses excluded from scans due to the 30 day unresponsive filter & \sm{638.6} & \sk{18.5} & 64.9 \\
		6Graph \cite{yang20226Graph} & Applied on responsive addresses in December 2021 & \sm{125.8} & \sk{18.9} & 65.2 \\
		6Tree \cite{liu20196Tree} & Applied on responsive addresses in December 2021 & \sm{37.6} & \sk{15.0} & 51.7 \\
		6GAN \cite{cui20216GAN} & Applied on responsive addresses in December 2021 & \sm{3.3} & 249 & 0.8 \\
		6VecLM \cite{cui20216VecLM} & Applied on responsive addresses in December 2021 & \sk{70.3} & 278 & 0.9 \\
		Distance clustering & Extending clustered address regions & \sm{5.3} & \sk{7.2} &  25.0\\
		\bottomrule
	\end{tabular}
\end{table*}

\paragraph{\textbf{Are other Protocols Responsive?}}
Based on the previous findings, we tested aliased prefixes for their responsiveness to all protocols.
We excluded \sk{66.4} aliased prefixes announced by Trafficforce, to reduce the impact as they were only responsive to ICMP probes during the initial multi-level aliased detection.
We only selected a single, random  address from each prefix to reduce traffic on real aliases and due to our assumption that all addresses behave the same.
Scan results for \sk{42.8} aliased prefixes can be seen in \Cref{tab:aliased_proto_scan}.
Most probed addresses were responsive to ICMP or TCP/80, and already tested during the multi-level aliased prefix detection.
Additionally, TCP/443 and especially UDP/443 were supported by a majority of tested addresses as well.
As shown in \Cref{tab:ip_as}, UDP/443 accounted for the lowest number of responsive addresses with \sk{98.1} in the \hitlist.
Therefore, using a single address from each aliased prefix increases the amount of responsive addresses by \sperc{29.4}.
This is in line with our observations that aliased prefixes are frequently seen in combination with \acp{cdn} and the findings by \citeauthor{zirngibl2021over9000} \cite{zirngibl2021over9000} that large providers mainly drive the deployment of QUIC.
In contrast, only 172 addresses were responsive to UDP/53 probes, \eg from Cloudflare or Misaka (AS50069), an anycast DNS service. 

In this scan, aliased prefix were responsive to at most four protocols. 
Only Cloudflare originates at least one prefix responsive to each probe respectively.
In no prefix was UDP/443 and UDP/53 seen in combination.

\subsection{\textit{Takeaways and Suggestions}}
The \emph{aliased prefix} detection of the \hitlist service is an important feature, necessary to allow an ongoing, feasible service and prevent biases in  the set of responsive  addresses, \eg towards Amazon with more than \sm{200} addresses.
However, the initial definition as an alias for a single host does not necessarily hold and the number of aliased prefixes is increasing over time.
Their complete removal can result in the exclusion of complete \acp{as}, \eg Fastly, or targets hosting multiple millions of domains, \eg within Cloudflare.
An informed assessment of these fully responsive prefixes depending on the usage of hitlist results is essential and can drastically change results and improve insights.

\textit{We suggest extending information regarding these address regions, \eg statistics about hosted domains, to allow an informed selection of address candidates out of these.
Furthermore, at least a single address out of each aliased prefix can be added to the hitlist.
Even if the complete prefix is an alias for a single host, testing its responsiveness with one of these addresses can cover its behavior and offer a valuable foundation for future research.
As shown in \Cref{tab:aliased_proto_scan}, random addresses are often responsive to different protocol scans.
However, known addresses contained in the input from DNS scans or passive sources should be used if possible because these addresses are either actively announced by operators in DNS or known to be used by network devices (\eg responsive to RIPE Atlas traceroutes).
}

\section{Discovering New Addresses}
\label{sec:new}
Besides the analysis of the historical state of the \hitlist and identified aliased prefixes, we analyze new potential address sources to further improve the service.
As shown in \Cref{fig:hitlist_pipeline}, we use the \hitlist service pipeline to filter and scan these addresses.

\begin{table*}[h]
	
	\begin{threeparttable}
		\caption{Responsive addresses for new sources divided by protocol. The top \acp{as} for each source based on the number of responsive addresses indicates potential biases in each data set. \acp{as} are abbreviated as symbols explained in the footer of the table. \hitlist results are from April 7, 2022.}
		\label{tab:responsiveness_new_sources}
		\begin{tabular}{lrrrrrrrcrcr}
			\toprule
			& \multicolumn{6}{c}{Responsive Addresses} & \multicolumn{5}{c}{ASes} \\
			\cmidrule(lr){2-7} \cmidrule(lr){8-12}
			Source &     {ICMP} &  TCP/443 &   TCP/80 &  UDP/443 &  UDP/53 & Total & \multicolumn{2}{c}{Top 1} & \multicolumn{2}{c}{Top 2}  & Total \\
			\midrule
			6Graph       			&  \sm{3.8} 	&  \sk{428.4} 	& \sk{491.1} &  \sk{121.9} &  \sk{78.6} & \sm{3.8} & \sperc{52.1} & $\blackdiamond$ & \sperc{5.1} & $\blacktriangledown$ & \sk{10.7}\\
			6Tree        			&  \sm{2.2} 	&  \sk{374.2} 	& \sk{425.5} &  \sk{116.6} &  \sk{62.8} & \sm{2.2} & \sperc{41.0} & $\blackdiamond$ & \sperc{8.0} & $\blacktriangledown$ & \sk{11.5} \\
			Unresponsive addresses 	&  \sm{1.2} 	&  \sk{274.8} 	& \sk{282.3} &   \sk{18.6} &  \sk{51.6} & \sm{1.3} & \sperc{34.4} & $\blacktriangle$ & \sperc{6.2} & $\bluetriangledown$ & \sk{9.0} \\
			Distance clustering     &  \sk{637.1} 	&  \sk{167.7} 	& \sk{193.4} &   \sk{85.1} & \sk{32.4} & \sk{651.0}  & \sperc{14.9} & $\bluebullet$ & \sperc{10.9} & $\bluestar$ & \sk{5.5}\\
			Passive sources      	&  \sk{21.0} 	&  \sk{1.5} 	& \sk{1.9} 	 &     358 &   3012 & \sk{21.6} & \sperc{6.7} & $\bullet$ & \sperc{3.2} & $\medstar$ & \sk{2.9} \\
			6GAN         			&  \sk{4.3}		&	      27 	&  			28	&       2 	 &      2 &  \sk{4.3} & \sperc{82.8} & $\bluediamond$ & \sperc{12.3} & $\bluetriangle$ & 39 \\
			6VecLM       			&      990 		&     103 &     116 &      38 &     22 & \sk{1.0} & \sperc{17.1} & $\bluediamond$ & \sperc{14.9} & $\reddiamond$ & 105\\
			\midrule
			New Sources 			& \sm{5.4} 		& \sk{764.9} & \sk{843.4} & \sk{164.0} & \sk{144.3} & \sm{5.6} & \sperc{26.8} & $\blackdiamond$ & \sperc{5.8} & $\blacktriangle$ & \sk{14.6} \\
			\hitlist 				& \sm{3.2} 		& \sk{910.8} & \sm{1.1} & \sk{98.1} &  \sk{140.7} & \sm{3.2} & \sperc{7.9} & $\redtriangle$ & \sperc{7.4} & $\redtriangledown$ & \sk{15.7} \\
			Total 					& \sm{8.6} 		& \sm{1.7} & \sm{1.9} & \sk{266.2} & \sk{287.4} & \sm{8.8} & \sperc{25.5} & $\blackdiamond$ & \sperc{5.5} & $\blacktriangle$ & \sk{17.3}\\
			\bottomrule
		\end{tabular}
		\begin{tablenotes}
			\item $\blackdiamond$ Free SAS (AS12322), $\blacktriangle$ VNPT (AS45899), $\blacktriangledown$ DigitalOcean (AS14061), $\bullet$ China Mobile (AS9808), $\medstar$ Racktech (AS208861),
			\item $\bluediamond$ CERN (AS513) $\bluetriangle$, ARNES (AS2107), $\bluetriangledown$ home.pl (AS12824), $\bluebullet$ Deutsche Glasfaser (AS60294), $\bluestar$ Akamai (AS20940),
			\item $\reddiamond$ Level3 (AS3356), $\redtriangle$ Linode (AS63949), $\redtriangledown$ China Telecom (AS4812)
		\end{tablenotes}
	\end{threeparttable}
\end{table*}

\subsection{New Sources}
The \hitlist was initially created with a variety of sources \cite{gasser2018clusters}.
However, other sources are available via different protocols (\eg MX and NS records) or scanning from different vantage points (\eg CAIDA Ark \cite{caidaArk}).
Furthermore, \citeauthor{gasser2018clusters} \cite{gasser2018clusters} showed the potential of target generation algorithms as part of the \hitlist input collection.
The reachability  of IPv6 addresses from available vantage points impacts the identification of  candidates, \eg due to location specific load balancing.
Additionally, active sources such as DNS scans and traceroutes are biased towards available scan targets.
They can only detect IPv6 addresses mapped to a domain or responsive to traceroutes.
Target generation algorithms try to mitigate these disadvantages from other sources.

We identified the following list of sources as potential candidates to improve the \hitlist and consolidate its value to the community.
An overview of all new sources can be seen in \Cref{tab:new_sources}.

\paragraph{\textbf{Passive Sources}}
Due to changes in the availability of data sources (\eg the discontinuation of the free availability of Rapid7 data), an ongoing update of the initial data collection efforts is necessary.
We collected a set of new passive sources, trying to identify additional responsive addresses that can be used as an extension of the existing service.
This includes results from our DNS scans as explained in \Cref{sec:data}.
While our institution actively conducts these scans, they were not specifically set up for this work and data was available for use.
Furthermore, we extract IPv6 addresses from the CAIDA Ark traceroute efforts and the published list of responsive addresses by \citeauthor{song2022det} \cite{song2022det}.
In total, these sources result in \sm{3.5} candidates.

However, \sperc{90} of these addresses were already contained in the service, \eg from its DNS resolutions or traceroutes, and \sperc{7.5} were additionally located in aliased prefixes.
\sk{369.1} (\sperc{71}) of the IPv6 addresses related to NS and MX records were located within Amazon, a highly aliased prefix.
Therefore, these sources result in only \sk{356.7} new addresses in total out of which, \sk{84.9} were not aliased.

\paragraph{\textbf{Unresponsive Addresses}}
As described in \Cref{sec:data}, the \hitlist service stops scanning addresses unresponsive for more than 30 days and never re-evaluates their responsiveness.
This list accumulates \sm{787.7} addresses.
We applied the blocklist filter and removed all candidates that showed \ac{gfw} injection, resulting in \sm{638.6} remaining candidates.
The explicit inclusion of these addresses into the scan requires deactivating the final filter of the \hitlist service (see \Cref{fig:hitlist_pipeline}).
We re-scanned these addresses once to get insights into whether addresses are responsive again and find that these should be re-evaluated regularly as shown in the next section.

\paragraph{\textbf{Target Generation}}

We applied different target generation algorithms to the set of responsive addresses of December 2021, including 6Graph \cite{yang20226Graph}, 6Tree \cite{liu20196Tree}, 6GAN \cite{cui20216GAN} and 6VecLM \cite{cui20216VecLM}.
We did not try to optimize or tune the algorithms but follow the respective explanations and standard parameters.
Furthermore, we generated addresses with a simple approach named distance clustering (DC), extending more densely clustered address regions that show high entropy in the last nibble(s) of the address.
Note, these regions were not fully responsive but only densely populated.
Therefore, we collected clusters of addresses with at least 10 addresses and a distance of at most 64 between two addresses.
Given the vast address space of IPv6, even a few addresses (10) within this comparably small distance are highly likely not assigned randomly but based on active assignment policies.
We generated missing addresses within these clusters.
We tested for potential new aliased prefixes after extending these clusters before our scans.

\Cref{tab:new_sources} lists how many addresses were generated by each applied method.
While the respective publications often limit the number of generated addresses, \eg to \sk{50}, we invest more computation time to increase the number of generated candidates.
However, we did not test different parameters or subsets of our input.
We used published code and the contained default configuration.

In theory, 6Tree actively scans candidates during target generation and uses results to improve detection.
To reduce scan impact, it contains functionality to detect aliases.
However, it did not detect aliased prefixes effectively in our initial tests, quickly inducing bias towards fully responsive regions.
6Tree generated a set of more than \sm{8.3} IPv6 addresses part of a single /48 prefix originated by Akamai (AS20940).
Most of these addresses were incrementally assigned and responsive but not labeled as aliased by the given 6Tree implementation.
However, the aliased prefix detection of the \hitlist identified these addresses as aliased.
Therefore, we prevented active scans, limited 6Tree to target generation only, and used the detection proposed by the \hitlist service during our scans.

\subsection{New Responsive Addresses}
We used ZMapv6 to scan all previously introduced, non-aliased addresses for ICMP, TCP/443, TCP/80, UDP/443, and UDP/53.
We scanned multiple times across four weeks to account for packet loss or network events and aggregated the results afterward.
The only source we did not completely scan multiple times is the set of unresponsive addresses due to its size and thus ethical reasons.
Here, we only included responsive addresses during the first test into following scans.

Even though we filtered responses injected by the \ac{gfw} before generating addresses, some algorithms generated many addresses located within Chinese \acp{as} and thus were affected by the \ac{gfw}.
Considering 6Graph, \sm{18.5} (\sperc{14.5}) out of the generated \sm{125.8} addresses were affected.
We filtered these responses for the following analysis.
\Cref{tab:responsiveness_new_sources} shows the cleaned number of responsive addresses for each source.

For the new passive sources, \sperc{25} of non-aliased addresses were responsive (\sk{21} out of \sk{84.9}).
For \sm{638.6} previously excluded addresses, unresponsive for at least 30 days, more than \sm{1.2} addresses were responsive again.
However, the responsiveness of these addresses decreased after the initial scan to nearly half the number of addresses.
We plan to regularly include a subset of these addresses into the existing ongoing service.
This allows testing addresses regularly but spreads the scan load over a longer period.

Regarding target generation algorithms, our naive approach to extend densely assigned address regions achieved better results than more sophisticated approaches, namely 6GAN and 6VecLM.
The latter methods only resulted in \sk{1} and \sk{4.4} responsive addresses respectively.
The hit rate of both algorithms was low even before filtering \ac{gfw} injected responses.
Our naive distance clustering approach resulted in \sk{651} addresses with a hit rate of nearly \sperc{12}.

6Tree and 6Graph resulted in the highest number of responsive addresses with \sm{2.2} and \sm{3.8}, respectively.
However, due to their large number of generated addresses, their discovery rate was \sperc{6} and \sperc{3}, respectively.

In total, we were able to identify and generate \sm{5.4} new or previously removed addresses responsive to at least one protocol.
This results in nearly twice as many responsive addresses for UDP/53 in total and three times more addresses for UDP/443.
Interestingly, slightly fewer addresses were responsive to TCP/80 and TCP/443 compared to the current \hitlist.
However, new sources identify \sperc{168} more responsive addresses to ICMP probes compared to the \hitlist. 
This is mostly due to the bias of target generation algorithms towards specific providers which we discuss next.

\begin{figure}
	\includegraphics[width=\linewidth]{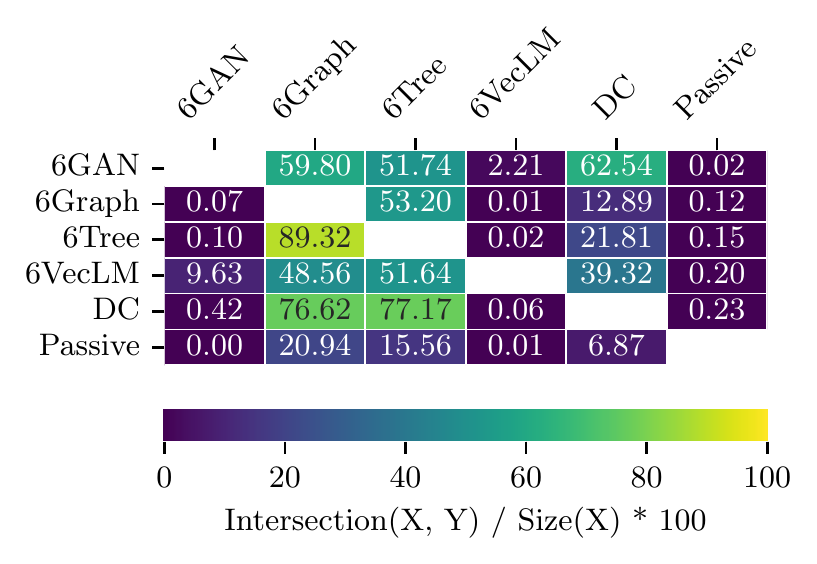}
	\caption{Overlap between responsive addresses from new sources in \% in respect to the total number of responsive addresses for each row.}
	\label{fig:overlap}
\end{figure}

\paragraph{\textbf{Overlap}}
Individual sources do not necessarily contribute unique responsive addresses to the \hitlist but an overlap between new sources is visible.
\Cref{fig:overlap} shows the overlap between all new sources in relation to the total number of responsive addresses for each row in percent.
Thus, \sperc{89.34} of all responsive addresses generated by 6Tree were also generated by 6Graph.
All target generation algorithms identified responsive addresses which were also part of passive sources and thus visible in traceroutes or \ac{dns} data.
Furthermore, all sources provided unique responsive addresses and thus show potential to improve the hitlist in the future. 
Even though new passive sources analyzed in this work only provide \sk{21.6} addresses (\sperc{0.7} of to the total of \sm{3.2} on April 7th, 2022), they offer new responsive candidates that are also not covered by target generation algorithms.
Especially with the end of freely available data sources, \eg Rapid7 forward DNS data, a frequent evaluation of new sources will be valuable.

\begin{figure}
	\includegraphics[width=\linewidth]{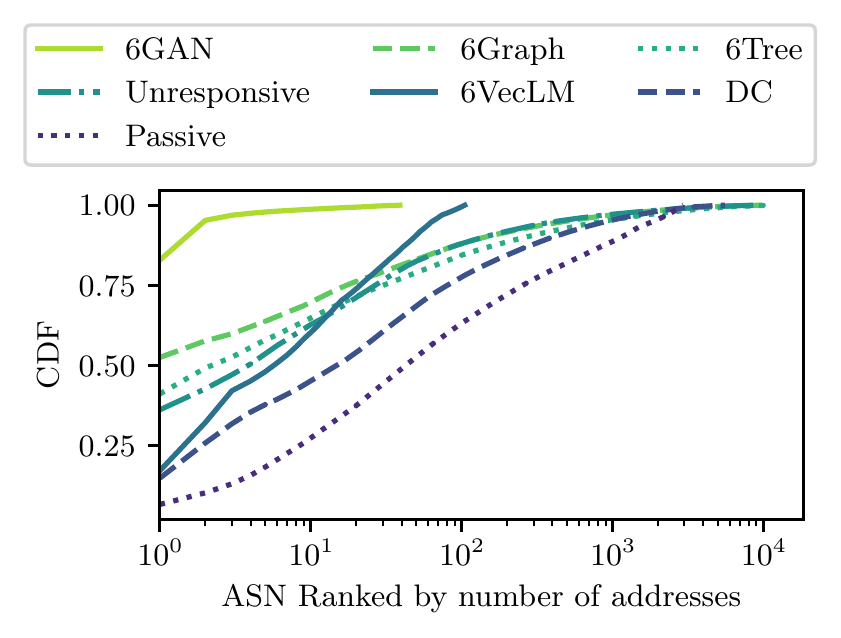}
	\caption{AS distribution of responsive addresses from new inputs.}
	\label{fig:new_inputs_asn}
\end{figure}

\paragraph{\textbf{Distribution Across ASes}}
To analyze whether new addresses impose a new bias to the \hitlist, we analyze their \acs{as} distribution.
\Cref{fig:new_inputs_asn} shows the distribution of responsive addresses across \acp{as} for each new input while the total number and most common \acp{as} are listed in \Cref{tab:responsiveness_new_sources}.
6Graph and 6Tree, contributing most new addresses, both show a visible bias towards Free SAS/ProXad covering up to \sperc{52} of the respective candidates.
The second most prominent \ac{as}, DigitalOcean, is only at around \sperc{5} to \sperc{8}.
We verified the correct classification of these addresses as non-aliased and came to the same conclusion as the automatic detection.
While many addresses were responsive, we can identify unresponsive addresses. 
The existing \hitlist already contains \sk{149.8} responsive addresses from Free SAS on April 7th, 2022.

Other sources provide different \ac{as} distributions and top hitters, \eg many previously unresponsive addresses were from VNPT.
The distance clustering approach and new passive sources show the most even distribution.
The latter even covers \sk{2.9} \acp{as} with only \sk{21} responsive addresses.

\subsection{\textit{Takeaways and Suggestions}}
\sperc{90} of passively extracted addresses from existing data sets were already contained in the hitlist or unique addresses which were however aliased.
In contrast, address generation approaches were able to identify new, previously unknown but responsive addresses.
While their individual hit rate was below \sperc{10}, combining all responsive addresses from passive sources and address generation can more than double the number of responsive addresses currently reported by the \hitlist.

Often, a major difficulty with these techniques is their incomplete documentation.
Hence, we were not able to reproduce results of 6GAN \cite{cui20216GAN} but it only generated \sk{4} responsive addresses.

\textit{Nevertheless, we suggest to regularly include addresses from some of these approaches and to include new candidates in the future.}

\section{Discussion}
\label{sec:discussion}
In this section, we summarize key-findings of our work and discuss results.

\paragraph{\textbf{Fully Responsive Prefixes}}
The \hitlist tries to identify \emph{aliased} prefixes, where a complete prefix is used by a single host.
However, we show that identified prefixes are not necessarily used by a single host, but these are sometimes a result of \acs{cdn} or middlebox deployments.
As shown in \Cref{sec:aliased}, multiple \acp{cdn} assign large parts of their owned address space to their fleet of servers, \eg Fastly or Cloudflare.
Amazon alone would introduce \sm{200} addresses from fully responsive prefixes.
These addresses still do not represent individual hosts each, introduce significant biases, and result in a massively increased scan load.
However, these prefixes are used in combination with multiple hosts for load balancing as seen with the \acl{tbt} and host multiple millions of domains including highly ranked domains according to top lists but also name servers or mail exchangers.
We suggest naming identified prefixes \emph{fully responsive prefixes} in the future and to analyze their characteristics in more detail to allow for better differentiation.
Fully responsive prefixes are a superset of aliased prefixes, identifiable by the implemented multi-level aliased prefix detection.
However, not all identified prefixes are actual aliases on a single machine.

While we will exclude most of these targets from the ongoing scans of the \hitlist, we argue that a subset of these addresses can be included and should be used by follow-up research.
Especially in the case of higher layer protocol scans and analyses, such as TLS or QUIC scans, these targets should not be excluded but considered, \eg based on up-to-date DNS resolutions.

We suggest that for each fully responsive prefix, at least a single address can be tested for responsiveness by the service.
Even if the prefix is an actual alias for a single host, it is an actual host, is part of the Internet ecosystem, and should thus be represented in the \hitlist and considered in the future.
Regarding fully responsive prefixes related to \acp{cdn}, we suggest a use case specific selection of addresses.
Single addresses can be enough to identify server setups, \eg in respect to TLS where a centrally administered configuration within a \ac{cdn} can be expected.
In contrast, multiple addresses might be required to analyze hosted domains, certificates or websites to spread the load, \eg for Cloudflare with multiple million domains within a single fully responsive prefix.

\paragraph{\textbf{New Input Sources}}
As shown in \Cref{sec:new}, updating the \hitlist with new input sources can be valuable to improve its quality and increase the amount of identified responsive addresses.
We collected and successfully applied a set of different target generation methods.
While we were not able to reproduce most hit rates published with target generation mechanisms, we can identify new responsive targets, more than doubling the number of responsive addresses.
However, a comprehensive and public evaluation is only possible if generated candidates and results are shared.
This would further improve the impact of these methodologies in general and on the \hitlist.

While target generation mechanisms sometimes bias towards certain providers, such as Free SaS, generated targets still cover a variety of \acp{as} and show a similar distribution across protocols as the current \hitlist.
Nevertheless, solely relying on those generated candidates shifts the focus towards different providers and deployments and changes the view on the IPv6 ecosystem.
In our opinion all sources (existing passive sources, frequent traceroutes, generated candidates) provide individual value, and their contribution to the \hitlist offers a valuable foundation.

We incorporated new sources into the ongoing \hitlist service, and plan to update the service with new sources on a regular basis.
A new target generation algorithm announced by \citeauthor{song2022addrminer} \cite{song2022addrminer} additionally proposes a methodology to identify potential address candidates in previously uncovered \acp{as}.
However, their results and tool were not yet accessible as of September  6th, 2022.
This research direction shows the potential to increase the coverage of announced prefixes by the \hitlist from currently \sperc{62} of announced IPv6 prefixes (see \Cref{sec:hitlist}).

\paragraph{\textbf{Service Maintenance}}
An intrinsic motivation of this research was to improve the foundation for IPv6 research and thus improve a valuable existing building block, namely the \hitlist service \cite{gasser2018clusters}.
Ongoing measurement studies that freely publish results and data are valuable research resources.
The \hitlist has been operated for more than four years and has been used  by a multitude of research 
\cite{zirngibl2021over9000,aschenbrenner2021mptcp,nosyk2022loops,almeida2020loadbalancing,deccio2020closeddoors,rodday2021defaultroutes,liu20196Tree,cui20206GCVAE,yang20226Graph,cui20216GAN,cui20216VecLM}.
However, its operation and maintenance requires continual effort which can be difficult especially in an academic environment.
While different approaches have evaluated the hitlist and proposed new address generation methodologies, they did not establish a new continuously running service.
We updated the service and included newly identified addresses into the service to improve the hitlist for future use.

To better enable future collaborations to maintain the \hitlist service and allow reproducible network measurement studies, we additionally share all data used for this work and used analysis scripts \cite{datatum}.
This includes our adaptation to 6Tree, the distance clustering implementation and a tool to filter \zmap output as published by the \hitlist service from the \ac{gfw} injection.

\section{Conclusion}
\label{sec:conclusion}
In this work, we took steps to dust the \hitlist service after four years of ongoing operation and prepare it for the following years.
Our steps include the analysis of the current state, cleaning the list of addresses caused by the \acl{gfw}, and evaluating the development of the \hitlist over time.
We further analyzed identified aliased prefixes and highlight that their strict omission excludes large \acp{cdn} such as Fastly or Cloudflare and thus multiple million domains from research based on the \hitlist.
Due to their load balancing mechanisms, announced prefixes appear to be aliased even though the backing infrastructure can be expected to be more than a single host and is valuable to analyze.
Finally, we evaluate different address candidate sources, their potentially induced biases and responsiveness, which can more than double the set of responsive addresses of the \hitlist.
The impact of our findings regarding the \ac{gfw} and new address sources is already visible in the current state of the hitlist and further steps are planned.

\label{body}

\begin{acks}
	The authors would like to thank the anonymous reviewers for their valuable feedback. This work
	was partially funded by the German Federal Ministry of Education
	and Research under the project PRIMEnet, grant 16KIS1370.
\end{acks}

\balance

\bibliographystyle{ACM-Reference-Format}
\bibliography{reference}


\begin{thebibliography}{52}

%%% the Web address.

\ifx \showCODEN    \undefined \def \showCODEN     #1{\unskip}     \fi
\ifx \showDOI      \undefined \def \showDOI       #1{#1}\fi
\ifx \showISBNx    \undefined \def \showISBNx     #1{\unskip}     \fi
\ifx \showISBNxiii \undefined \def \showISBNxiii  #1{\unskip}     \fi
\ifx \showISSN     \undefined \def \showISSN      #1{\unskip}     \fi
\ifx \showLCCN     \undefined \def \showLCCN      #1{\unskip}     \fi
\ifx \shownote     \undefined \def \shownote      #1{#1}          \fi
\ifx \showarticletitle \undefined \def \showarticletitle #1{#1}   \fi
\ifx \showURL      \undefined \def \showURL       {\relax}        \fi
% invisible to TeX
\providecommand\bibfield[2]{#2}
\providecommand\bibinfo[2]{#2}
\providecommand\natexlab[1]{#1}
\providecommand\showeprint[2][]{arXiv:#2}

\bibitem[\protect\citeauthoryear{{Alexa}}{{Alexa}}{2022}]%
        {alexa}
\bibfield{author}{\bibinfo{person}{{Alexa}}.} \bibinfo{year}{2022}\natexlab{}.
\newblock \bibinfo{booktitle}{\emph{{Top 1M sites}}}.
\newblock
\urldef\tempurl%
\url{https://www.alexa.com/topsites}
\showURL{%
\tempurl}


\bibitem[\protect\citeauthoryear{Almeida, Cunha, Teixeira, Veitch, and
  Diot}{Almeida et~al\mbox{.}}{2020}]%
        {almeida2020loadbalancing}
\bibfield{author}{\bibinfo{person}{Rafael Almeida}, \bibinfo{person}{ítalo
  Cunha}, \bibinfo{person}{Renata Teixeira}, \bibinfo{person}{Darryl Veitch},
  {and} \bibinfo{person}{Christophe Diot}.} \bibinfo{year}{2020}\natexlab{}.
\newblock \showarticletitle{{Classification of Load Balancing in the
  Internet}}. In \bibinfo{booktitle}{\emph{Proc. IEEE Int. Conference on
  Computer Communications (INFOCOM)}}.
\newblock


\bibitem[\protect\citeauthoryear{Alt, Beverly, and Dainotti}{Alt
  et~al\mbox{.}}{2014}]%
        {alt2014v4aliases}
\bibfield{author}{\bibinfo{person}{Lance Alt}, \bibinfo{person}{Robert
  Beverly}, {and} \bibinfo{person}{Alberto Dainotti}.}
  \bibinfo{year}{2014}\natexlab{}.
\newblock \showarticletitle{{Uncovering Network Tarpits with Degreaser}}. In
  \bibinfo{booktitle}{\emph{Proceedings of the 30th Annual Computer Security
  Applications Conference}} (New Orleans, Louisiana, USA).
\newblock


\bibitem[\protect\citeauthoryear{Anonymous}{Anonymous}{2014}]%
        {anonymous2014comprehensivefirewallpicture}
\bibfield{author}{\bibinfo{person}{Anonymous}.}
  \bibinfo{year}{2014}\natexlab{}.
\newblock \showarticletitle{{Towards a Comprehensive Picture of the Great
  Firewall{\textquoteright}s DNS Censorship}}. In \bibinfo{booktitle}{\emph{4th
  USENIX Workshop on Free and Open Communications on the Internet (FOCI 14)}}.
  \bibinfo{publisher}{USENIX Association}, \bibinfo{address}{San Diego, CA}.
\newblock
\urldef\tempurl%
\url{https://www.usenix.org/conference/foci14/workshop-program/presentation/anonymous}
\showURL{%
\tempurl}


\bibitem[\protect\citeauthoryear{Anonymous, Niaki, Hoang, Gill, and
  Houmansadr}{Anonymous et~al\mbox{.}}{2020}]%
        {anonymous2020tripletcensors}
\bibfield{author}{\bibinfo{person}{Anonymous}, \bibinfo{person}{Arian~Akhavan
  Niaki}, \bibinfo{person}{Nguyen~Phong Hoang}, \bibinfo{person}{Phillipa
  Gill}, {and} \bibinfo{person}{Amir Houmansadr}.}
  \bibinfo{year}{2020}\natexlab{}.
\newblock \showarticletitle{{Triplet Censors: Demystifying Great
  Firewall{\textquoteright}s DNS Censorship Behavior}}. In
  \bibinfo{booktitle}{\emph{10th USENIX Workshop on Free and Open
  Communications on the Internet (FOCI 20)}}. \bibinfo{publisher}{USENIX
  Association}.
\newblock
\urldef\tempurl%
\url{https://www.usenix.org/conference/foci20/presentation/anonymous}
\showURL{%
\tempurl}


\bibitem[\protect\citeauthoryear{APNIC}{APNIC}{2022}]%
        {apnicv6stats}
\bibfield{author}{\bibinfo{person}{APNIC}.} \bibinfo{year}{2022}\natexlab{}.
\newblock \bibinfo{booktitle}{\emph{{IPv6 Capable Rate by country (\%)}}}.
\newblock
\urldef\tempurl%
\url{https://stats.labs.apnic.net/ipv6}
\showURL{%
Retrieved 2022-05-07 from \tempurl}


\bibitem[\protect\citeauthoryear{Aschenbrenner, Shreedhar, Gasser, Mohan, and
  Ott}{Aschenbrenner et~al\mbox{.}}{2021}]%
        {aschenbrenner2021mptcp}
\bibfield{author}{\bibinfo{person}{Florian Aschenbrenner},
  \bibinfo{person}{Tanya Shreedhar}, \bibinfo{person}{Oliver Gasser},
  \bibinfo{person}{Nitinder Mohan}, {and} \bibinfo{person}{Jörg Ott}.}
  \bibinfo{year}{2021}\natexlab{}.
\newblock \showarticletitle{{From Single Lane to Highways: Analyzing the
  Adoption of Multipath TCP in the Internet}}. In
  \bibinfo{booktitle}{\emph{2021 IFIP Networking Conference (IFIP
  Networking)}}.
\newblock


\bibitem[\protect\citeauthoryear{Backes, Holz, Rossow, Rytilahti, Simeonovski,
  and Stock}{Backes et~al\mbox{.}}{2016}]%
        {backes2016ittl}
\bibfield{author}{\bibinfo{person}{Michael Backes}, \bibinfo{person}{Thorsten
  Holz}, \bibinfo{person}{Christian Rossow}, \bibinfo{person}{Teemu Rytilahti},
  \bibinfo{person}{Milivoj Simeonovski}, {and} \bibinfo{person}{Ben Stock}.}
  \bibinfo{year}{2016}\natexlab{}.
\newblock \showarticletitle{{On the Feasibility of TTL-Based Filtering for
  DRDoS Mitigation}}. In \bibinfo{booktitle}{\emph{Research in Attacks,
  Intrusions, and Defenses}}.
\newblock


\bibitem[\protect\citeauthoryear{Beverly}{Beverly}{2016}]%
        {beverly2016yarrp}
\bibfield{author}{\bibinfo{person}{Robert Beverly}.}
  \bibinfo{year}{2016}\natexlab{}.
\newblock \showarticletitle{{Yarrp'ing the Internet: Randomized High-Speed
  Active Topology Discovery}}. In \bibinfo{booktitle}{\emph{Proc. ACM Int.
  Measurement Conference (IMC)}} (Santa Monica, California, USA).
\newblock


\bibitem[\protect\citeauthoryear{Beverly, Brinkmeyer, Luckie, and
  Rohrer}{Beverly et~al\mbox{.}}{2013}]%
        {beverly2013tbt}
\bibfield{author}{\bibinfo{person}{Robert Beverly}, \bibinfo{person}{William
  Brinkmeyer}, \bibinfo{person}{Matthew Luckie}, {and}
  \bibinfo{person}{Justin~P. Rohrer}.} \bibinfo{year}{2013}\natexlab{}.
\newblock \showarticletitle{{IPv6 Alias Resolution via Induced Fragmentation}}.
  In \bibinfo{booktitle}{\emph{Proc. Passive and Active Measurement (PAM)}}.
\newblock


\bibitem[\protect\citeauthoryear{{CAIDA}}{{CAIDA}}{2022}]%
        {caidaArk}
\bibfield{author}{\bibinfo{person}{{CAIDA}}.} \bibinfo{year}{2022}\natexlab{}.
\newblock \bibinfo{booktitle}{\emph{{Ark IPv6 Topology Dataset}}}.
\newblock
\urldef\tempurl%
\url{https://catalog.caida.org/details/dataset/ipv6_allpref_topology}
\showURL{%
Retrieved 2022-05-11 from \tempurl}


\bibitem[\protect\citeauthoryear{{Cisco}}{{Cisco}}{2022}]%
        {umbrella}
\bibfield{author}{\bibinfo{person}{{Cisco}}.} \bibinfo{year}{2022}\natexlab{}.
\newblock \bibinfo{booktitle}{\emph{{Umbrella Top 1M List}}}.
\newblock
\urldef\tempurl%
\url{https://umbrella.cisco.com/blog/cisco-umbrella-1-million}
\showURL{%
\tempurl}


\bibitem[\protect\citeauthoryear{Cui, Gou, and Xiong}{Cui
  et~al\mbox{.}}{2020}]%
        {cui20206GCVAE}
\bibfield{author}{\bibinfo{person}{Tianyu Cui}, \bibinfo{person}{Gaopeng Gou},
  {and} \bibinfo{person}{Gang Xiong}.} \bibinfo{year}{2020}\natexlab{}.
\newblock \showarticletitle{{6GCVAE: Gated Convolutional Variational
  Autoencoder for IPv6 Target Generation}}. In
  \bibinfo{booktitle}{\emph{Advances in Knowledge Discovery and Data Mining}}.
  \bibinfo{publisher}{Springer International Publishing}.
\newblock


\bibitem[\protect\citeauthoryear{Cui, Gou, Xiong, Liu, Fu, and Li}{Cui
  et~al\mbox{.}}{2021a}]%
        {cui20216GAN}
\bibfield{author}{\bibinfo{person}{Tianyu Cui}, \bibinfo{person}{Gaopeng Gou},
  \bibinfo{person}{Gang Xiong}, \bibinfo{person}{Chang Liu},
  \bibinfo{person}{Peipei Fu}, {and} \bibinfo{person}{Zhen Li}.}
  \bibinfo{year}{2021}\natexlab{a}.
\newblock \showarticletitle{{6GAN: IPv6 Multi-Pattern Target Generation via
  Generative Adversarial Nets with Reinforcement Learning}}. In
  \bibinfo{booktitle}{\emph{Proc. IEEE Int. Conference on Computer
  Communications (INFOCOM)}}.
\newblock


\bibitem[\protect\citeauthoryear{Cui, Xiong, Gou, Shi, and Xia}{Cui
  et~al\mbox{.}}{2021b}]%
        {cui20216VecLM}
\bibfield{author}{\bibinfo{person}{Tianyu Cui}, \bibinfo{person}{Gang Xiong},
  \bibinfo{person}{Gaopeng Gou}, \bibinfo{person}{Junzheng Shi}, {and}
  \bibinfo{person}{Wei Xia}.} \bibinfo{year}{2021}\natexlab{b}.
\newblock \showarticletitle{{6VecLM: Language Modeling in Vector Space for IPv6
  Target Generation}}. In \bibinfo{booktitle}{\emph{Machine Learning and
  Knowledge Discovery in Databases: Applied Data Science Track}}.
\newblock


\bibitem[\protect\citeauthoryear{Deccio, Hilton, Briggs, Avery, and
  Richardson}{Deccio et~al\mbox{.}}{2020}]%
        {deccio2020closeddoors}
\bibfield{author}{\bibinfo{person}{Casey Deccio}, \bibinfo{person}{Alden
  Hilton}, \bibinfo{person}{Michael Briggs}, \bibinfo{person}{Trevin Avery},
  {and} \bibinfo{person}{Robert Richardson}.} \bibinfo{year}{2020}\natexlab{}.
\newblock \showarticletitle{{Behind Closed Doors: A Network Tale of Spoofing,
  Intrusion, and False DNS Security}}. In \bibinfo{booktitle}{\emph{Proc. ACM
  Int. Measurement Conference (IMC)}} (Virtual Event, USA).
\newblock


\bibitem[\protect\citeauthoryear{Dittrich, Kenneally, et~al\mbox{.}}{Dittrich
  et~al\mbox{.}}{2012}]%
        {menloreport}
\bibfield{author}{\bibinfo{person}{David Dittrich}, \bibinfo{person}{Erin
  Kenneally}, {et~al\mbox{.}}} \bibinfo{year}{2012}\natexlab{}.
\newblock \showarticletitle{{The Menlo Report: Ethical principles guiding
  information and communication technology research}}.
\newblock \bibinfo{journal}{\emph{US Department of Homeland Security}}
  (\bibinfo{year}{2012}).
\newblock


\bibitem[\protect\citeauthoryear{Farnan, Darer, and Wright}{Farnan
  et~al\mbox{.}}{2016}]%
        {farnan2016poisoningthewell}
\bibfield{author}{\bibinfo{person}{Oliver Farnan}, \bibinfo{person}{Alexander
  Darer}, {and} \bibinfo{person}{Joss Wright}.}
  \bibinfo{year}{2016}\natexlab{}.
\newblock \showarticletitle{{Poisoning the Well: Exploring the Great Firewall's
  Poisoned DNS Responses}}. In \bibinfo{booktitle}{\emph{Proceedings of the
  2016 ACM on Workshop on Privacy in the Electronic Society}} (Vienna,
  Austria).
\newblock


\bibitem[\protect\citeauthoryear{Fiebig, Borgolte, Hao, Kruegel, and
  Vigna}{Fiebig et~al\mbox{.}}{2017}]%
        {fiebig2018somethingfromnothing}
\bibfield{author}{\bibinfo{person}{Tobias Fiebig}, \bibinfo{person}{Kevin
  Borgolte}, \bibinfo{person}{Shuang Hao}, \bibinfo{person}{Christopher
  Kruegel}, {and} \bibinfo{person}{Giovanni Vigna}.}
  \bibinfo{year}{2017}\natexlab{}.
\newblock \showarticletitle{{Something from Nothing (There): Collecting Global
  IPv6 Datasets from DNS}}. In \bibinfo{booktitle}{\emph{Proc. Passive and
  Active Measurement (PAM)}}.
\newblock


\bibitem[\protect\citeauthoryear{Foremski, Plonka, and Berger}{Foremski
  et~al\mbox{.}}{2016}]%
        {foremski2016entropyip}
\bibfield{author}{\bibinfo{person}{Pawel Foremski}, \bibinfo{person}{David
  Plonka}, {and} \bibinfo{person}{Arthur Berger}.}
  \bibinfo{year}{2016}\natexlab{}.
\newblock \showarticletitle{{Entropy/IP: Uncovering Structure in IPv6
  Addresses}}. In \bibinfo{booktitle}{\emph{Proc. ACM Int. Measurement
  Conference (IMC)}} (Santa Monica, California, USA).
\newblock


\bibitem[\protect\citeauthoryear{Gasser}{Gasser}{2022}]%
        {zmapv6}
\bibfield{author}{\bibinfo{person}{Oliver Gasser}.}
  \bibinfo{year}{2022}\natexlab{}.
\newblock \bibinfo{booktitle}{\emph{{ZMapv6: Internet Scanner with IPv6
  capabilities}}}.
\newblock
\urldef\tempurl%
\url{https://github.com/tumi8/zmap}
\showURL{%
\tempurl}


\bibitem[\protect\citeauthoryear{Gasser, Scheitle, Foremski, Lone, Korczynski,
  Strowes, Hendriks, and Carle}{Gasser et~al\mbox{.}}{2018}]%
        {gasser2018clusters}
\bibfield{author}{\bibinfo{person}{Oliver Gasser}, \bibinfo{person}{Quirin
  Scheitle}, \bibinfo{person}{Pawel Foremski}, \bibinfo{person}{Qasim Lone},
  \bibinfo{person}{Maciej Korczynski}, \bibinfo{person}{Stephen~D. Strowes},
  \bibinfo{person}{Luuk Hendriks}, {and} \bibinfo{person}{Georg Carle}.}
  \bibinfo{year}{2018}\natexlab{}.
\newblock \showarticletitle{{Clusters in the Expanse: Understanding and
  Unbiasing IPv6 Hitlists}}. In \bibinfo{booktitle}{\emph{Proc. ACM Int.
  Measurement Conference (IMC)}} (Boston, MA, USA).
\newblock


\bibitem[\protect\citeauthoryear{Gasser, Scheitle, Gebhard, and Carle}{Gasser
  et~al\mbox{.}}{2016}]%
        {gasser2016scanning}
\bibfield{author}{\bibinfo{person}{Oliver Gasser}, \bibinfo{person}{Quirin
  Scheitle}, \bibinfo{person}{Sebastian Gebhard}, {and} \bibinfo{person}{Georg
  Carle}.} \bibinfo{year}{2016}\natexlab{}.
\newblock \showarticletitle{{Scanning the IPv6 Internet: Towards a
  Comprehensive Hitlist}}. In \bibinfo{booktitle}{\emph{Proc. 8th Int. Workshop
  on Traffic Monitoring and Analysis}}. \bibinfo{address}{Louvain-la-Neuve,
  Belgium}.
\newblock


\bibitem[\protect\citeauthoryear{Google}{Google}{2022}]%
        {googlev6growth}
\bibfield{author}{\bibinfo{person}{Google}.} \bibinfo{year}{2022}\natexlab{}.
\newblock \bibinfo{booktitle}{\emph{{Statistics: IPv6 Adoption}}}.
\newblock
\urldef\tempurl%
\url{https://www.google.com/intl/en/ipv6/statistics.html}
\showURL{%
Retrieved 2022-05-07 from \tempurl}


\bibitem[\protect\citeauthoryear{Hou, Cai, Wu, Su, and Xiong}{Hou
  et~al\mbox{.}}{2021}]%
        {hou20216Hit}
\bibfield{author}{\bibinfo{person}{Bingnan Hou}, \bibinfo{person}{Zhiping Cai},
  \bibinfo{person}{Kui Wu}, \bibinfo{person}{Jinshu Su}, {and}
  \bibinfo{person}{Yinqiao Xiong}.} \bibinfo{year}{2021}\natexlab{}.
\newblock \showarticletitle{{6Hit: A Reinforcement Learning-based Approach to
  Target Generation for Internet-wide IPv6 Scanning}}. In
  \bibinfo{booktitle}{\emph{Proc. IEEE Int. Conference on Computer
  Communications (INFOCOM)}}.
\newblock


\bibitem[\protect\citeauthoryear{Huitema}{Huitema}{2006}]%
        {rfc4380}
\bibfield{author}{\bibinfo{person}{Christian Huitema}.}
  \bibinfo{year}{2006}\natexlab{}.
\newblock \bibinfo{title}{{Teredo: Tunneling IPv6 over UDP through Network
  Address Translations (NATs)}}.
\newblock \bibinfo{howpublished}{RFC 4380}.
\newblock


\bibitem[\protect\citeauthoryear{Iyengar and Thomson}{Iyengar and
  Thomson}{2021}]%
        {rfc9000}
\bibfield{author}{\bibinfo{person}{Jana Iyengar} {and} \bibinfo{person}{Martin
  Thomson}.} \bibinfo{year}{2021}\natexlab{}.
\newblock \bibinfo{title}{{QUIC: A UDP-Based Multiplexed and Secure
  Transport}}.
\newblock \bibinfo{howpublished}{RFC 9000}.
\newblock
\urldef\tempurl%
\url{https://doi.org/10.17487/RFC9000}
\showDOI{\tempurl}


\bibitem[\protect\citeauthoryear{Izhikevich, Teixeira, and
  Durumeric}{Izhikevich et~al\mbox{.}}{2021}]%
        {izhikevich2021lzr}
\bibfield{author}{\bibinfo{person}{Liz Izhikevich}, \bibinfo{person}{Renata
  Teixeira}, {and} \bibinfo{person}{Zakir Durumeric}.}
  \bibinfo{year}{2021}\natexlab{}.
\newblock \showarticletitle{{LZR}: Identifying Unexpected Internet Services}.
  In \bibinfo{booktitle}{\emph{Proc. USENIX Security Symposium}}.
\newblock
\urldef\tempurl%
\url{https://www.usenix.org/conference/usenixsecurity21/presentation/izhikevich}
\showURL{%
\tempurl}


\bibitem[\protect\citeauthoryear{Liu, Xiong, Liu, Xie, and Zhu}{Liu
  et~al\mbox{.}}{2019}]%
        {liu20196Tree}
\bibfield{author}{\bibinfo{person}{Zhizhu Liu}, \bibinfo{person}{Yinqiao
  Xiong}, \bibinfo{person}{Xin Liu}, \bibinfo{person}{Wei Xie}, {and}
  \bibinfo{person}{Peidong Zhu}.} \bibinfo{year}{2019}\natexlab{}.
\newblock \showarticletitle{{6Tree: Efficient dynamic discovery of active
  addresses in the IPv6 address space}}.
\newblock \bibinfo{journal}{\emph{Computer Networks}} (\bibinfo{year}{2019}).
\newblock
\urldef\tempurl%
\url{https://www.sciencedirect.com/science/article/pii/S1389128618312003}
\showURL{%
\tempurl}


\bibitem[\protect\citeauthoryear{Luckie, Beverly, Brinkmeyer, and
  claffy}{Luckie et~al\mbox{.}}{2013}]%
        {luckie2013speedtrap}
\bibfield{author}{\bibinfo{person}{Matthew Luckie}, \bibinfo{person}{Robert
  Beverly}, \bibinfo{person}{William Brinkmeyer}, {and} \bibinfo{person}{kc
  claffy}.} \bibinfo{year}{2013}\natexlab{}.
\newblock \showarticletitle{{Speedtrap: Internet-Scale IPv6 Alias Resolution}}.
  In \bibinfo{booktitle}{\emph{Proc. ACM Int. Measurement Conference (IMC)}}
  (Barcelona, Spain).
\newblock


\bibitem[\protect\citeauthoryear{{Majestic}}{{Majestic}}{2022}]%
        {majestic}
\bibfield{author}{\bibinfo{person}{{Majestic}}.}
  \bibinfo{year}{2022}\natexlab{}.
\newblock \bibinfo{booktitle}{\emph{{The Majestic Million}}}.
\newblock
\urldef\tempurl%
\url{https://majestic.com/reports/majestic-million/}
\showURL{%
\tempurl}


\bibitem[\protect\citeauthoryear{Marder}{Marder}{2020}]%
        {marder2020apple}
\bibfield{author}{\bibinfo{person}{Alexander Marder}.}
  \bibinfo{year}{2020}\natexlab{}.
\newblock \showarticletitle{{APPLE: Alias Pruning by Path Length Estimation}}.
  In \bibinfo{booktitle}{\emph{Proc. Passive and Active Measurement (PAM)}}.
\newblock


\bibitem[\protect\citeauthoryear{Maxmind}{Maxmind}{2022}]%
        {maxmind}
\bibfield{author}{\bibinfo{person}{Maxmind}.} \bibinfo{year}{2022}\natexlab{}.
\newblock \bibinfo{booktitle}{\emph{{GeoLite2 Free Geolocation Data}}}.
\newblock
\urldef\tempurl%
\url{https://dev.maxmind.com/geoip/geolite2-free-geolocation-data?lang=en}
\showURL{%
\tempurl}


\bibitem[\protect\citeauthoryear{Mukaddam, Elhajj, Kayssi, and Chehab}{Mukaddam
  et~al\mbox{.}}{2014}]%
        {mukaddam2014ittl}
\bibfield{author}{\bibinfo{person}{Ayman Mukaddam}, \bibinfo{person}{Imad
  Elhajj}, \bibinfo{person}{Ayman Kayssi}, {and} \bibinfo{person}{Ali Chehab}.}
  \bibinfo{year}{2014}\natexlab{}.
\newblock \showarticletitle{{IP Spoofing Detection Using Modified Hop Count}}.
  In \bibinfo{booktitle}{\emph{2014 IEEE 28th International Conference on
  Advanced Information Networking and Applications}}.
\newblock


\bibitem[\protect\citeauthoryear{Murdock, Li, Bramsen, Durumeric, and
  Paxson}{Murdock et~al\mbox{.}}{2017}]%
        {murdock20176Gen}
\bibfield{author}{\bibinfo{person}{Austin Murdock}, \bibinfo{person}{Frank Li},
  \bibinfo{person}{Paul Bramsen}, \bibinfo{person}{Zakir Durumeric}, {and}
  \bibinfo{person}{Vern Paxson}.} \bibinfo{year}{2017}\natexlab{}.
\newblock \showarticletitle{{Target Generation for Internet-Wide IPv6
  Scanning}}. In \bibinfo{booktitle}{\emph{Proc. ACM Int. Measurement
  Conference (IMC)}} (London, United Kingdom).
\newblock


\bibitem[\protect\citeauthoryear{NCC}{NCC}{2022a}]%
        {riperis}
\bibfield{author}{\bibinfo{person}{RIPE NCC}.}
  \bibinfo{year}{2022}\natexlab{a}.
\newblock \bibinfo{booktitle}{\emph{{Routing Information Service (RIS)}}}.
\newblock
\urldef\tempurl%
\url{https://www.ripe.net/analyse/internet-measurements/routing-information-service-ris}
\showURL{%
Retrieved 2022-08-29 from \tempurl}


\bibitem[\protect\citeauthoryear{NCC}{NCC}{2022b}]%
        {ripev6growth}
\bibfield{author}{\bibinfo{person}{RIPE NCC}.}
  \bibinfo{year}{2022}\natexlab{b}.
\newblock \bibinfo{booktitle}{\emph{{Total IPv6 Allocations and Assignments}}}.
\newblock
\urldef\tempurl%
\url{https://www.ripe.net/analyse/statistics/total-ipv6-allocations_and_assignments}
\showURL{%
Retrieved 2022-05-07 from \tempurl}


\bibitem[\protect\citeauthoryear{Nosyk, Korczy{\'{n}}ski, and Duda}{Nosyk
  et~al\mbox{.}}{2022}]%
        {nosyk2022loops}
\bibfield{author}{\bibinfo{person}{Yevheniya Nosyk}, \bibinfo{person}{Maciej
  Korczy{\'{n}}ski}, {and} \bibinfo{person}{Andrzej Duda}.}
  \bibinfo{year}{2022}\natexlab{}.
\newblock \showarticletitle{{Routing Loops as Mega Amplifiers for DNS-Based
  DDoS Attacks}}. In \bibinfo{booktitle}{\emph{Proc. Passive and Active
  Measurement (PAM)}}.
\newblock


\bibitem[\protect\citeauthoryear{Padmanabhan, Li, Levin, and
  Spring}{Padmanabhan et~al\mbox{.}}{2015}]%
        {padmanabhan2015uav6}
\bibfield{author}{\bibinfo{person}{Ramakrishna Padmanabhan},
  \bibinfo{person}{Zhihao Li}, \bibinfo{person}{Dave Levin}, {and}
  \bibinfo{person}{Neil Spring}.} \bibinfo{year}{2015}\natexlab{}.
\newblock \showarticletitle{{UAv6: Alias Resolution in IPv6 Using Unused
  Addresses}}. In \bibinfo{booktitle}{\emph{Proc. Passive and Active
  Measurement (PAM)}}.
\newblock


\bibitem[\protect\citeauthoryear{Partridge and Allman}{Partridge and
  Allman}{2016}]%
        {PA16}
\bibfield{author}{\bibinfo{person}{Craig Partridge} {and} \bibinfo{person}{Mark
  Allman}.} \bibinfo{year}{2016}\natexlab{}.
\newblock \showarticletitle{{{Addressing Ethical Considerations in Network
  Measurement Papers}}}.
\newblock \bibinfo{journal}{\emph{Commun. ACM}} \bibinfo{volume}{59},
  \bibinfo{number}{10} (\bibinfo{date}{Oct.} \bibinfo{year}{2016}).
\newblock


\bibitem[\protect\citeauthoryear{Poese, Uhlig, Kaafar, Donnet, and Gueye}{Poese
  et~al\mbox{.}}{2011}]%
        {poese2011geolocation}
\bibfield{author}{\bibinfo{person}{Ingmar Poese}, \bibinfo{person}{Steve
  Uhlig}, \bibinfo{person}{Mohamed~Ali Kaafar}, \bibinfo{person}{Benoit
  Donnet}, {and} \bibinfo{person}{Bamba Gueye}.}
  \bibinfo{year}{2011}\natexlab{}.
\newblock \showarticletitle{{IP Geolocation Databases: Unreliable?}}
\newblock \bibinfo{journal}{\emph{ACM SIGCOMM Computer Communication Review}}
  (\bibinfo{year}{2011}).
\newblock


\bibitem[\protect\citeauthoryear{Rodday, Kaltenbach, Cunha, Bush, Katz-Bassett,
  Rodosek, Schmidt, and W\"{a}hlisch}{Rodday et~al\mbox{.}}{2021}]%
        {rodday2021defaultroutes}
\bibfield{author}{\bibinfo{person}{Nils Rodday}, \bibinfo{person}{Lukas
  Kaltenbach}, \bibinfo{person}{Italo Cunha}, \bibinfo{person}{Randy Bush},
  \bibinfo{person}{Ethan Katz-Bassett}, \bibinfo{person}{Gabi~Dreo Rodosek},
  \bibinfo{person}{Thomas~C. Schmidt}, {and} \bibinfo{person}{Matthias
  W\"{a}hlisch}.} \bibinfo{year}{2021}\natexlab{}.
\newblock \showarticletitle{{On the Deployment of Default Routes in
  Inter-Domain Routing}} \emph{(\bibinfo{series}{TAURIN'21})}.
\newblock


\bibitem[\protect\citeauthoryear{Rye, Beverly, and Claffy}{Rye
  et~al\mbox{.}}{2021}]%
        {rye2021followthescent}
\bibfield{author}{\bibinfo{person}{Erik Rye}, \bibinfo{person}{Robert Beverly},
  {and} \bibinfo{person}{K~C Claffy}.} \bibinfo{year}{2021}\natexlab{}.
\newblock \showarticletitle{{Follow the Scent: Defeating IPv6 Prefix Rotation
  Privacy}}. In \bibinfo{booktitle}{\emph{Proc. ACM Int. Measurement Conference
  (IMC)}} (Virtual Event).
\newblock


\bibitem[\protect\citeauthoryear{Scheitle, Gasser, Rouhi, and Carle}{Scheitle
  et~al\mbox{.}}{2017a}]%
        {scheitle2017clockskew}
\bibfield{author}{\bibinfo{person}{Quirin Scheitle}, \bibinfo{person}{Oliver
  Gasser}, \bibinfo{person}{Minoo Rouhi}, {and} \bibinfo{person}{Georg Carle}.}
  \bibinfo{year}{2017}\natexlab{a}.
\newblock \showarticletitle{{Large-scale classification of IPv6-IPv4 siblings
  with variable clock skew}}. In \bibinfo{booktitle}{\emph{Proc. Network
  Traffic Measurement and Analysis Conference (TMA)}}.
\newblock


\bibitem[\protect\citeauthoryear{Scheitle, Gasser, Sattler, and Carle}{Scheitle
  et~al\mbox{.}}{2017b}]%
        {scheitle2017hloc}
\bibfield{author}{\bibinfo{person}{Quirin Scheitle}, \bibinfo{person}{Oliver
  Gasser}, \bibinfo{person}{Patrick Sattler}, {and} \bibinfo{person}{Georg
  Carle}.} \bibinfo{year}{2017}\natexlab{b}.
\newblock \showarticletitle{{HLOC: Hints-Based Geolocation Leveraging Multiple
  Measurement Frameworks}}. In \bibinfo{booktitle}{\emph{Proc. Network Traffic
  Measurement and Analysis Conference (TMA)}}.
\newblock


\bibitem[\protect\citeauthoryear{Song, Yang, He, Wang, Li, Duan, Liu, and
  Sun}{Song et~al\mbox{.}}{2022a}]%
        {song2022addrminer}
\bibfield{author}{\bibinfo{person}{Guanglei Song}, \bibinfo{person}{Jiahai
  Yang}, \bibinfo{person}{Lin He}, \bibinfo{person}{Zhiliang Wang},
  \bibinfo{person}{Guo Li}, \bibinfo{person}{Chenxin Duan},
  \bibinfo{person}{Yaozhong Liu}, {and} \bibinfo{person}{Zhongxiang Sun}.}
  \bibinfo{year}{2022}\natexlab{a}.
\newblock \showarticletitle{{AddrMiner: A Comprehensive Global Active IPv6
  Address Discovery System}}. In \bibinfo{booktitle}{\emph{2022 USENIX Annual
  Technical Conference (USENIX ATC 22)}}. \bibinfo{address}{Carlsbad, CA}.
\newblock
\urldef\tempurl%
\url{https://www.usenix.org/conference/atc22/presentation/song}
\showURL{%
\tempurl}


\bibitem[\protect\citeauthoryear{Song, Yang, Wang, He, Lin, Pan, Duan, and
  Quan}{Song et~al\mbox{.}}{2022b}]%
        {song2022det}
\bibfield{author}{\bibinfo{person}{Guanglei Song}, \bibinfo{person}{Jiahai
  Yang}, \bibinfo{person}{Zhiliang Wang}, \bibinfo{person}{Lin He},
  \bibinfo{person}{Jinlei Lin}, \bibinfo{person}{Long Pan},
  \bibinfo{person}{Chenxin Duan}, {and} \bibinfo{person}{Xiaowen Quan}.}
  \bibinfo{year}{2022}\natexlab{b}.
\newblock \showarticletitle{{DET: Enabling Efficient Probing of IPv6 Active
  Addresses}}.
\newblock \bibinfo{journal}{\emph{IEEE/ACM Transactions on Networking}}
  (\bibinfo{year}{2022}).
\newblock


\bibitem[\protect\citeauthoryear{Vermeulen, Ljuma, Addanki, Gouel, Fourmaux,
  Friedman, and Rejaie}{Vermeulen et~al\mbox{.}}{2020}]%
        {vermeulen2020ratelimiting}
\bibfield{author}{\bibinfo{person}{Kevin Vermeulen}, \bibinfo{person}{Burim
  Ljuma}, \bibinfo{person}{Vamsi Addanki}, \bibinfo{person}{Matthieu Gouel},
  \bibinfo{person}{Olivier Fourmaux}, \bibinfo{person}{Timur Friedman}, {and}
  \bibinfo{person}{Reza Rejaie}.} \bibinfo{year}{2020}\natexlab{}.
\newblock \showarticletitle{{Alias Resolution Based on ICMP Rate Limiting}}. In
  \bibinfo{booktitle}{\emph{Proc. Passive and Active Measurement (PAM)}}.
\newblock


\bibitem[\protect\citeauthoryear{Yang, Hou, Cai, Wu, Zhou, and Wang}{Yang
  et~al\mbox{.}}{2022}]%
        {yang20226Graph}
\bibfield{author}{\bibinfo{person}{Tao Yang}, \bibinfo{person}{Bingnan Hou},
  \bibinfo{person}{Zhiping Cai}, \bibinfo{person}{Kui Wu},
  \bibinfo{person}{Tongqing Zhou}, {and} \bibinfo{person}{Chengyu Wang}.}
  \bibinfo{year}{2022}\natexlab{}.
\newblock \showarticletitle{{6Graph: A graph-theoretic approach to address
  pattern mining for Internet-wide IPv6 scanning}}.
\newblock \bibinfo{journal}{\emph{Computer Networks}} (\bibinfo{year}{2022}).
\newblock
\urldef\tempurl%
\url{https://www.sciencedirect.com/science/article/pii/S1389128621005430}
\showURL{%
\tempurl}


\bibitem[\protect\citeauthoryear{Zhang, Lu, Liu, Duan, and Liu}{Zhang
  et~al\mbox{.}}{2022}]%
        {zhang2022DNSRootserver}
\bibfield{author}{\bibinfo{person}{Fenglu Zhang}, \bibinfo{person}{Chaoyi Lu},
  \bibinfo{person}{Baojun Liu}, \bibinfo{person}{Haixin Duan}, {and}
  \bibinfo{person}{Ying Liu}.} \bibinfo{year}{2022}\natexlab{}.
\newblock \showarticletitle{{Measuring the Practical Effect of DNS Root Server
  Instances: A China-Wide Case Study}}. In \bibinfo{booktitle}{\emph{Proc.
  Passive and Active Measurement (PAM)}}.
\newblock


\bibitem[\protect\citeauthoryear{Zirngibl, Buschmann, Sattler, Jaeger, Aulbach,
  and Carle}{Zirngibl et~al\mbox{.}}{2021}]%
        {zirngibl2021over9000}
\bibfield{author}{\bibinfo{person}{Johannes Zirngibl},
  \bibinfo{person}{Philippe Buschmann}, \bibinfo{person}{Patrick Sattler},
  \bibinfo{person}{Benedikt Jaeger}, \bibinfo{person}{Juliane Aulbach}, {and}
  \bibinfo{person}{Georg Carle}.} \bibinfo{year}{2021}\natexlab{}.
\newblock \showarticletitle{{It's over 9000: Analyzing early QUIC Deployments
  with the Standardization on the Horizon}}. In \bibinfo{booktitle}{\emph{Proc.
  ACM Int. Measurement Conference (IMC)}} (Virtual Event, USA).
\newblock


\bibitem[\protect\citeauthoryear{Zirngibl, Steger, Sattler, Gasser, and
  Carle}{Zirngibl et~al\mbox{.}}{2022}]%
        {datatum}
\bibfield{author}{\bibinfo{person}{Johannes Zirngibl}, \bibinfo{person}{Lion
  Steger}, \bibinfo{person}{Patrick Sattler}, \bibinfo{person}{Oliver Gasser},
  {and} \bibinfo{person}{Georg Carle}.} \bibinfo{year}{2022}\natexlab{}.
\newblock \bibinfo{booktitle}{\emph{{Data and Analysis at TUM University
  Library}}}.
\newblock
\urldef\tempurl%
\url{https://mediatum.ub.tum.de/1686542}
\showURL{%
\tempurl}
\newblock
\shownote{doi:10.14459/2022mp1686542}.


\end{thebibliography}
\appendix
\section{Autonomous Systems Impacted by the GFW}
\label{app:as}
\Cref{tab:appendix_china_as} shows the top 10 \acp{as} accounting for addresses in the \hitlist impacted by the \acl{gfw}.
The effect and impact on the \hitlist is explained in detail in \Cref{sec:hitlist}.
The \hitlist vantage point is located in Germany and thus, probes sent into Chinese \acp{as}, crossing the \ac{gfw} are impacted.
The general distribution of impacted addresses across \acp{as} can be seen in \Cref{fig:complete_asn_cdf}.

\begin{table}[hb]
	\caption{Top 10 ASes of addresses impacted by the GFW. The total accumulated number of impacted addresses is 134~M.}
	\label{tab:appendix_china_as}
	\begin{tabular}{lrrr}
		\toprule
		ASes &  \# Addresses &       \% & CDF \\
		\midrule
		4134 &   \sm{62.3} &  46.44 &  46.44 \\
		4812 &   \sm{19.5} &  14.59 &  61.03 \\
		134774 &   \sm{18.6} &  13.88 & 74.92 \\
		134773 &   \sm{10.7} &  8.04 &  82.96 \\
		140329 &    \sm{3.1} &  2.37 &  85.34 \\
		134772 &    \sm{2.5} &  1.93 &  87.28 \\
		4837 &    \sm{2.5} &  1.87 &  89.17 \\
		136200 &    \sm{2.3} &  1.76 &  90.94 \\
		140330 &    \sm{2.3} &  1.72 &  92.66 \\
		140316 &    \sm{1.6} &  1.24 &  93.91 \\
		\bottomrule
	\end{tabular}
\end{table}

\section{Additional Information about Responsive Addresses per Protocol}
\label{app:responsive}

\begin{figure}[hb]
	\includegraphics{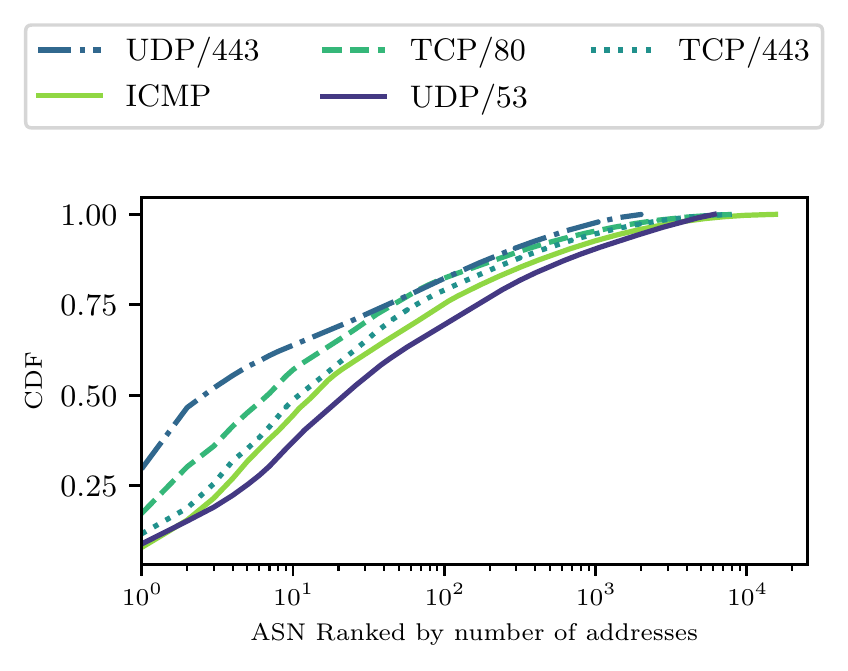}
	\caption{AS distribution of addresses responsive to each protocol on April 7th, 2022}
	\label{fig:asn_resp_proto}
\end{figure}

\begin{figure}[hb]
	\includegraphics{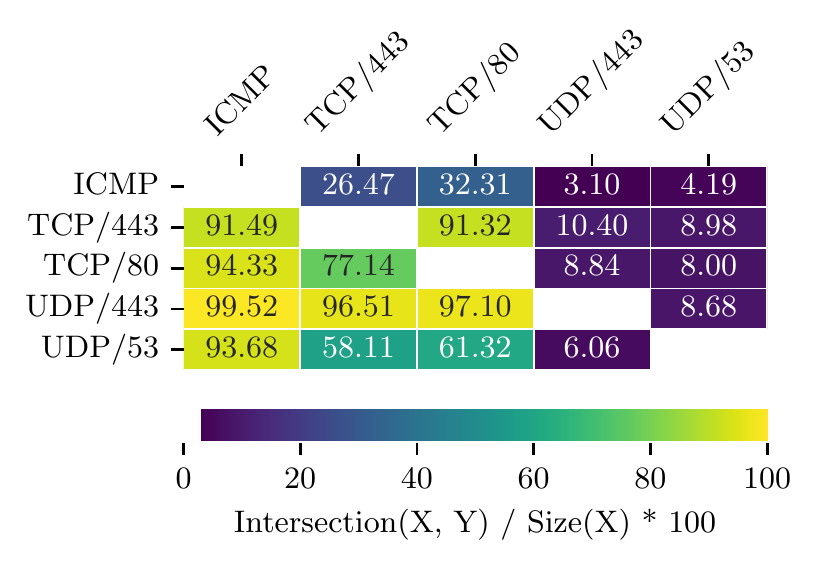}
	\caption{Overlap of addresses responsive to each protocol in the \hitlist on April 7th, 2022}
	\label{fig:resp_overlap}
\end{figure}

\Cref{fig:asn_resp_proto} and \ref{fig:resp_overlap} extend the results from \Cref{sec:hitlist} with further details regarding probed protocols.
\Cref{fig:asn_resp_proto} shows the \ac{as} distribution of responsive addresses from the \hitlist for each protocol as extension to \Cref{fig:complete_asn_cdf}.
Addresses responsive to UDP/53 show the most even distribution while UDP/443 is limited to the smallest number of different \acp{as}.
\Cref{fig:resp_overlap} shows the overlap between addresses responsive to each protocol.
Addresses responsive to TCP and UDP on each port are mostly also responsive to ICMP.
Additionally, large overlaps can be seen between TCP/80 (HTTP), TCP/443 (HTTPS) and UDP/443 (QUIC, HTTP3).

\label{lastpage}

\end{document}